\documentclass[dvips]{article}
\usepackage{epsfig}
\usepackage{amssymb}
\usepackage{amsmath}
\usepackage{subfigure}
\usepackage{pslatex}

\textheight=22cm \DeclareSymbolFont{ppa}{OT1}{ppl}{m}{it}
\DeclareMathSymbol{\vv}{\mathalpha}{ppa}{'166}

minus 2.3mu \thickmuskip = 2.6mu plus 2mu minus 2.6mu

\begin{document}

\newcommand{\dd}{\,{\rm d}}
\newcommand{\ie}{{\it i.e.},\,}
\newcommand{\etal}{{\it et al.\ }}
\newcommand{\eg}{{\it e.g.},\,}
\newcommand{\cf}{{\it cf.\ }}
\newcommand{\vs}{{\it vs.\ }}
\newcommand{\zdot}{\makebox[0pt][l]{.}}
\newcommand{\up}[1]{\ifmmode^{\rm #1}\else$^{\rm #1}$\fi}
\newcommand{\dn}[1]{\ifmmode_{\rm #1}\else$_{\rm #1}$\fi}
\newcommand{\upd}{\up{d}}
\newcommand{\uph}{\up{h}}
\newcommand{\upm}{\up{m}}
\newcommand{\ups}{\up{s}}
\newcommand{\arcd}{\ifmmode^{\circ}\else$^{\circ}$\fi}
\newcommand{\arcm}{\ifmmode{'}\else$'$\fi}
\newcommand{\arcs}{\ifmmode{''}\else$''$\fi}
\newcommand{\MS}{{\rm M}\ifmmode_{\odot}\else$_{\odot}$\fi}
\newcommand{\RS}{{\rm R}\ifmmode_{\odot}\else$_{\odot}$\fi}
\newcommand{\LS}{{\rm L}\ifmmode_{\odot}\else$_{\odot}$\fi}

\newcommand{\Abstract}[2]{{\vspace{1mm}\footnotesize\begin{center}ABSTRACT\end{center}
\vspace{1mm}\par#1\par \noindent {\bf Key words:~~}{\it #2}}}

\newcommand{\TabCap}[2]{\begin{center}\parbox[t]{#1}{\begin{center}
  \small {\spaceskip 2pt plus 1pt minus 1pt T a b l e}
  \refstepcounter{table}\thetable \\[2mm]
  \footnotesize #2 \end{center}}\end{center}}

\newcommand{\TableSep}[2]{\begin{table}[p]\vspace{#1}
\TabCap{#2}\end{table}}

\newcommand{\FigCap}[1]{\footnotesize\par\noindent Fig.\  %
  \refstepcounter{figure}\thefigure. #1\par}

\newcommand{\TableFont}{\footnotesize}
\newcommand{\TableFontIt}{\ttit}
\newcommand{\SetTableFont}[1]{\renewcommand{\TableFont}{#1}}

\newcommand{\MakeTable}[4]{\begin{table}[htb]\TabCap{#2}{#3}
  \begin{center} \TableFont \begin{tabular}{#1} #4
  \end{tabular}\end{center}\end{table}}

\newcommand{\MakeTableSep}[4]{\begin{table}[p]\TabCap{#2}{#3}
  \begin{center} \TableFont \begin{tabular}{#1} #4
  \end{tabular}\end{center}\end{table}}

\newenvironment{references}%
{ \footnotesize \frenchspacing
\renewcommand{\thesection}{}
\renewcommand{\in}{{\rm in }}
\renewcommand{\AA}{Astron.\ Astrophys.}
\newcommand{\AAS}{Astron.~Astrophys.~Suppl.~Ser.}
\newcommand{\AN}{Astron.~Nachr.}
\newcommand{\ApJ}{Astrophys.\ J.}
\newcommand{\ApJS}{Astrophys.\ J.~Suppl.~Ser.}
\newcommand{\ApJL}{Astrophys.\ J.~Letters}
\newcommand{\AJ}{Astron.\ J.}
\newcommand{\IBVS}{IBVS}
\newcommand{\PASP}{P.A.S.P.}
\newcommand{\Acta}{Acta Astron.}
\newcommand{\MNRAS}{MNRAS}
\renewcommand{\and}{{\rm and }}
\section{{\rm REFERENCES}}
\sloppy \hyphenpenalty10000
\begin{list}{}{\leftmargin1cm\listparindent-1cm
\itemindent\listparindent\parsep0pt\itemsep0pt}}%
{\end{list}\vspace{2mm}}

\def\TYLDA{~}
\newlength{\DW}
\settowidth{\DW}{0}
\newcommand{\dw}{\hspace{\DW}}

\newcommand{\refitem}[5]{\item[]{#1} #2%
\def\REFARG{#3}\ifx\REFARG\TYLDA\else, {\it#3}\fi
\def\REFARG{#4}\ifx\REFARG\TYLDA\else, {\bf#4}\fi
\def\REFARG{#5}\ifx\REFARG\TYLDA\else, {#5}\fi.}

\newcommand{\Section}[1]{\section{#1}}
\newcommand{\Subsection}[1]{\subsection{#1}}
\newcommand{\Acknow}[1]{\par\vspace{5mm}{\bf Acknowledgements.} #1}
\pagestyle{myheadings}

\newfont{\bb}{ptmbi8t at 12pt}
\newcommand{\xrule}{\rule{0pt}{2.5ex}}
\newcommand{\xxrule}{\rule[-1.8ex]{0pt}{4.5ex}}
\def\thefootnote{\fnsymbol{footnote}}

\begin{center}

{\Large\bf Black Holes Admitting Strong Resonant Phenomena}
\vskip0.5cm {\large
Z.~~S~t~u~c~h~l~\'{\i}~k,~~A.~~K~o~t~r~l~o~v~\'{a}~~and~~G.~~T~\"{o}~r~\"{o}~k}
\vskip5mm {Institute of Physics, Faculty of Philosophy and
Science, Silesian University in Opava, Bezru\v{c}ovo n\'{a}m. 13,
CZ-74601 Opava, Czech Republic\\
e-mail: andrea.kotrlova@fpf.slu.cz}
\end{center}

\Abstract{High-frequency twin peak quasiperiodic oscillations
(QPOs) are observed in four microquasars, \ie Ga\-lactic black
hole binary systems, with frequency ratio very close to 3\,:\,2.
In the microquasar GRS 1915+105, the structure of QPOs exhibits
additional frequencies, and more than two frequencies are observed
in the Galaxy nuclei Sgr\,A$^*$, or in some extragalactic sources
(NGC 4051, MCG-6-30-15 and NGC 5408 X-1). The observed QPOs can be
explained by a variety of the orbital resonance model versions
assuming resonance of oscillations with the Keplerian frequency
$\nu_{\mathrm{K}}$ or the vertical epicyclic frequency
$\nu_{\theta}$, and the radial epicyclic frequency
$\nu_{\mathrm{r}}$, or some combinations of these frequencies.
Generally, different resonances could arise at different radii of
an accretion disc. However, we have shown that for special values
of dimensionless black hole spin $a$ strong resonant phenomena
could occur when different resonances can be excited at the same
radius, as cooperative phenomena between the resonances may work
in such situations. The special values of $a$ are determined for
triple frequency ratio sets $\nu_{\mathrm{K}} : \nu_{\theta} :
\nu_{\mathrm{r}} = s:t:u$ with $s,t,u$ being small integers. The
most promising example of such a special situation arises for
black holes with extraordinary resonant spin $a=0.983$ at the
radius $r = 2.395\,M$, where
$\nu_{\mathrm{K}}$\,:\,$\nu_{\theta}$\,:\,$\nu_{\mathrm{r}} = $
3\,:\,2\,:\,1. We also predict that when combinations of the
orbital frequencies are allowed, QPOs with four frequency ratio
set 4\,:\,3\,:\,2\,:\,1 could be observed in the field of black
holes with $a = 0.866,\,0.882$ and $0.962$. Assuming the
extraordinary resonant spin $a=0.983$ in Sgr\,A$^*$, its QPOs with
observed frequency ratio $\approx$ 3\,:\,2\,:\,1 imply the black
hole mass in the interval $4.3 \times 10^6 ~\mathrm{M}_{\odot} < M
< 5.4 \times 10^6~\mathrm{M}_{\odot}$, in agreement with estimates
given by other, independent, observations.}{Accretion, accretion
disks -- Black hole physics -- X-rays: general}


\Section{Introduction} Quasiperiodic oscillations (\emph{QPOs}) of
X-ray brightness had been observed in many Galactic low-mass X-ray
binaries containing neutron~stars (see, \eg van der Klis 2000,
2006, Barret \etal 2005, Belloni \etal 2005, 2007) or black holes
(see, \eg McClintock and Remillard 2004, Remillard 2005, Remillard
and McClintock 2006). Some of the QPOs are in the kHz range and
often come in pairs $(\nu_{\rm upp}, \nu_{\rm down})$ of {\it twin
peaks} (often called \emph{double peaks}) in the Fourier power
spectra. Since the peaks of high frequencies are close to the
orbital frequency of the marginally stable circular orbit
representing the inner edge of Keplerian discs orbiting black
holes (or neutron~stars), the strong gra\-vi\-ty effects have to
be relevant in explaining high-frequency QPOs (Abramowicz \etal
2004).

The twin peak QPOs were observed in four microquasars, namely GRO
1655-40, XTE 1550-564, H 1743-322, GRS 1915+105 (T{\"{o}}r{\"{o}}k
\etal 2005). In all of the four cases, the frequency ratio of the
twin peaks is very close to 3\,:\,2 and the orbital resonance
model assuming non-linear resonances between oscillations in
Keplerian $\nu_{\mathrm{K}}$ and epicyclic frequencies
$\nu_{\theta}$ (vertical), or $\nu_{\mathrm{r}}$ (radial), or
their combinations, seems to be the potential explanation of the
microquasar kHz QPOs. The resonances of oscillations with the
epicyclic frequencies were firstly mentioned in Aliev and Galtsov
(1981). In the context of QPOs observed in the neutron star and
black hole systems, the orbital resonance model was introduced and
developed by Klu{\'z}niak and Abramowicz (2000), see Abramowicz
\etal (2004).

According to the resonance hypothesis, the two modes in resonance
should have eigenfrequencies $\nu_{\rm r}$ (equal to the radial
epicyclic frequency) and $\nu_{\rm v}$ (equal to the vertical
epicyclic frequency $\nu_{\rm{\theta}}$ or to the Keplerian
frequency $\nu_{\rm K}$). While models based on the
\emph{parametric resonance} identify the two observed frequencies
of the twin peaks ($\nu_\mathrm{upp}$, $\nu_\mathrm{down}$)
directly with the eigenfrequencies of a resonance, models based on
the \emph{forced resonance} allow to observe combinational (beat)
frequencies of the modes, or oscillations with the combinational
frequencies entering the resonance (Landau and Lifshitz 1976).
Both parametric and forced resonance models make clear and precise
predictions about the values of observed frequencies in connection
with spin and mass of the observed object, at least in the case of
black holes (T{\"{o}}r{\"{o}}k \etal 2005).

In some sources, more than two high-frequency peaks are observed.
The microquasar GRS 1915+105 reveals high-frequency QPOs appearing
at four frequencies with the lower and upper pairs in the ratio
close to 3\,:\,2 (Remillard and McClintock 2006) and even a fifth
frequency was reported (Belloni \etal 2001). An additional sixth
frequency was mentioned (Strohmayer 2001), although not confirmed.
In Sgr\,A$^*$, three frequencies were reported with ratio close to
3\,:\,2\,:\,1 (Aschenbach 2004, Aschenbach \etal 2004,
T{\"{o}}r{\"{o}}k 2005). In the galactic nuclei MCG-6-30-15 and
NGC 4051, two pairs of QPOs were reported with the ratios close to
3\,:\,2 and 2\,:\,1, respectively (Lachowicz \etal 2006). In the
source NGC 5408 X-1 oscillations with three frequencies of ratio
close to 6\,:\,4\,:\,3 were observed (Strohmayer \etal 2007).

In the microquasar GRS1915+105, an almost extreme Kerr black hole
with $a \approx 1$ is expected (McClintock \etal 2006), and all
the five (six) frequencies of QPOs can be explained in the
framework of the extended resonance model with the hump-induced
oscillations, predicting the black hole spin $a=0.9998$ and its
mass $M=14.8~\mathrm{M}_{\odot}$ (Stuchl{\'{\i}}k \etal 2006,
2007d). In the extended resonance model, forced resonances of the
epicyclic oscillations with an additional oscillation induced by
the humpy orbital velocity profile (related to the physically
privileged locally non-rotating frames) that occurs in Keplerian
discs orbiting Kerr black holes with $a > 0.9953$ are assumed
(Aschenbach 2004, 2006, Stuchl{\'{\i}}k \etal 2004, 2005, 2007c).
In the ``humpy'' extended resonance model, all the oscillations in
resonance can occur at, and be related to the exclusively defined,
``humpy radius'' with extremal orbital velocity gradient within
the humpy profile (Stuchl{\'{\i}}k \etal 2007c). However, this
model can be relevant only for near-extreme Kerr black holes with
spin $a > 0.9953$ for Keplerian discs, and with even higher spin
$a > 0.9998$ for marginally stable thick accretion discs
(Stuchl{\'{\i}}k \etal 2004, 2005).

In order to explain complex frequency structures observed in some
black hole systems, we expect that more than one resonance occur
in a Keplerian disc, and different versions of the orbital
resonance model could be acting. Of course, one version of the
orbital resonance model appearing at different resonant points, at
different radii of the disc, is also possible (Stuchl{\'{\i}}k and
T{\"{o}}r{\"{o}}k 2005). Note that the so called total precession
resonance model, assuming resonance of oscillations with the
Keplerian frequency $\nu_{\mathrm{K}}$ and the total precession
frequency $\nu_{\mathrm{T}}=\nu_{\theta}-\nu_{\mathrm{r}}$, can
explain quite well the multi-resonant phenomena observed in the
neutron star atoll source 4U 1636-53 (Stuchl{\'{\i}}k \etal 2007e)
and in some other atoll sources (Bakala \etal 2008, in
preparation).

Generally, two resonances could occur at different radii, and for
special values of the dimensionless spin $a$, the bottom, top, or
mixed frequencies coincide, reducing two frequency pairs into a
triple frequency set (Stuchl{\'{\i}}k \etal 2007b). Physically
more interesting seems to be the case, when a triple frequency set
$\nu_{\mathrm{K}}$, $\nu_{\theta}$, $\nu_{\mathrm{r}}$ with
rational ratio arises at a given radius. The resonant phenomena
can be then expected stronger than in the case of internally
independent resonances occurring at two different radii. Clearly,
while sharing a given radius, the resonances could be causally
related and could cooperate efficiently (Landau and Lifshitz
1976). That is the reason why we focus here attention on the
possibility of causally related resonances appearing at a fixed
radius. Such a situation is possible only in the field of Kerr
black holes with special values of the dimensionless spin $a$,
when the value of $a$ is related to the corresponding triple
frequency ratio set, and concrete versions of the resonance model
that are realized. Since the strength of the resonance and the
resonant frequency width decrease rapidly with the order of the
resonance $n+m$ (see Landau and Lifshitz 1976), in the following
we restrict ourselves to the cases with $n$, $m \leq 5$.

\Section{Orbital Resonance Model}
The standard orbital resonance
model (Abramowicz \etal 2004, T{\"{o}}r{\"{o}}k \etal 2005)
assumes oscillations of an accretion disc orbiting a rotating
black hole described by the Kerr geometry, or a neutron star that
could be represented by the Schwarzschild geometry, or, more
precisely, by the Hartle--Thorne geometry (Hartle and Thorne
1968). The accretion disc can be approximated by a thin disc with
Keplerian angular velocity profile, or by a thick toroidal disc
with angular velocity profile given by distribution of the
specific angular momentum in the fluid of the toroidal disc. The
frequency of the disc oscillations is related to the Keplerian
frequency (orbital frequency of tori), or the radial and vertical
epicyclic frequencies of the circular test particle (geodetical)
motion. The epicyclic frequencies can be relevant both for the
thin, Keplerian discs with quasicircular geodetical motion (Kato
\etal 1998, Novikov and Thorne 1973) and for thick, toroidal discs
with non-geodetical quasicircular rotation kept by the pressure
gradients of the tori (Schnittman and Rezzolla 2006, Rezzolla
\etal 2003, Rezzolla 2004ab). However, with thickness of an
oscillating toroid growing, the eigenfrequencies of its radial and
vertical oscillations deviate from the epicyclic test particle
frequencies ({\v S}r{\'{a}}mkov{\'{a}} 2005). Here we focus our
attention to the Keplerian thin discs.

Different versions of the orbital resonance model could be
classified according to the following criteria:
\begin{enumerate}
    \item[a)] the type of the resonance (parametric or forced),
    \item[b)] the type of oscillations entering the resonance,
    \item[c)] the presence of beat, combinational frequencies.
\end{enumerate}
Thus, according to the first criterion, two main groups of orbital
resonance model versions exist, differing by the type of the
resonance. In both of them, the epicyclic frequencies of the
equatorial test particle circular motion have a crucial role
(T{\"{o}}r{\"{o}}k \etal 2005).

The internal resonance model is based on the idea of
\emph{parametric resonance} between vertical and radial epicyclic
oscillations with the frequencies $\nu_\theta=\omega_ \theta/2\pi$
and $\nu_\mathrm{r}=\omega_\mathrm{r}/2\pi$. The parametric
resonance is described by the Mathieu equation (Landau and
Lifshitz 1976)
\begin{equation}
\label{Mathieu} \delta \ddot \theta + \omega_{\theta}^2\,\left[ 1
+ h \cos (\omega_\mathrm{r} t) \right]\, \delta \theta = 0.
\end{equation}

Theory behind the Mathieu equation implies that a parametric
resonance is excited when
\begin{equation}
\label{Equation6} {\frac{\omega_\mathrm{r}} {\omega_{\theta}}} =
{\frac{\nu_{\mathrm{r}}}  {\nu_{\theta}}} = {\frac{2}  {n}},
\qquad n =1, \,2, \,3,\, \dots
\end{equation}
and is strongest for the smallest possible value of $n$ (Landau
and Lifshitz 1976). Because $\nu_{\mathrm{r}} < \nu_{\theta}$ near
black holes, the smallest possible value for the parametric
resonance is $~n = 3$, which means that $2\, \nu_{\theta} = 3\,
\nu_\mathrm r$. This explains why the 3\,:\,2 ratio is commonly
observed in the black hole systems, assuming $\nu_{\rm upp} =
\nu_{\theta}$ and $\nu_{\rm down} = \nu_{\mathrm{r}}$. Note that
for the internal resonance the oscillating system conserves energy
(Landau and Lifshitz 1976).
\medskip

Versions of the resonance model based on the \emph{forced
resonance} come from the idea of a forced non-linear oscillator,
when the relation of the latitudinal (vertical) and radial
oscillations is given by the formulae
\begin{equation}
\delta \ddot \theta + \omega_{\theta}^2\delta \theta +
\left[{\mbox{non-linear~terms~in}}~\delta \theta \right] = g
(r)\cos (\omega_0\,t),
\end{equation}
\begin{equation}
\delta \ddot r + \omega_{\mathrm{r}}^2\delta r +
\left[{\mbox{non-linear~terms~in}}~\delta \theta, \delta r \right]
= h (r)\cos (\omega_0\,t),
\end{equation}
with
\begin{equation}
\omega_{\theta} = \left(\frac{p}{q}\right)\, \omega_\mathrm r,
\end{equation}
where $p,q$ are small natural numbers and $\omega_0$ is the
frequency of the external force, \textit{e.g.}, the gravitational
perturbative forces are discussed, for the case of a neutron star
with ``mountains'' or accretion columns, and a binary partner of
the neutron star or a black hole, in Stuchl{\'{\i}}k \etal (2008,
2007a), Stuchl{\'{\i}}k and Hled{\'{\i}}k (2005). The non-linear
terms allow the presence of combination (beat) frequencies in
resonant solutions for $\delta \theta (t)$ and $\delta r(t)$ (see,
\eg Landau and Lifshitz 1976), which in the simplest case give
\begin{equation}
\omega_- = \omega_{\theta} - \omega_\mathrm r, ~~~\omega_+
=\omega_{\theta} + \omega_\mathrm r.
\end{equation}

Another, so called \emph{``Keplerian'' resonance} model, takes
into account possible parametric or forced resonances between
oscillations with the radial epicyclic frequency $\nu_\mathrm{r}$
or the vertical epicyclic frequency $\nu_\mathrm{\theta}$, and the
Keplerian orbital frequency $\nu_\mathrm{K}$.
\medskip

Such resonances can produce the observable frequencies in the
3\,:\,2 ratio as well as in other rational ratii (note that one of
the cases which gives 3\,:\,2 observed ratio is also the
``direct'' case of $p:q=3:2$ corresponding to the same frequencies
and radius as in the case of 3\,:\,2 parametric resonance).
Therefore, we shall consider both the direct and simple
combinational resonances.

The resonance conditions of the parametric and direct forced
resonance are common, however, physical details, as the time
evolution of the resonance, the strength of the resonance and the
width of the resonant frequencies, are different (see Landau and
Lifshitz 1976).

The width of the resonant frequencies in forced resonances differs
in the cases of direct, sub(super)harmonic and combinational
resonances, and depends on the external force strength and the
damping and non-linear terms. Considering a simple case of
external force of harmonic character with frequency $\Omega$
influencing a non-linear oscillator with a single degree of
freedom and variable $u$, having eigenfrequency $\omega_0$,
damping parameter $\mu$ and cubic non-linear term characterized by
parameter $\alpha$, the equation of motion reads (Nayfeh and Mook
1979)
\begin{equation}
\ddot{u}+\omega_0^2\,u =
-2\,\epsilon\,\mu\,\dot{u}-\epsilon\,\alpha\,u^3 + \epsilon\,k
\cos \left(\Omega t\right)
\end{equation}
where $\epsilon$ is a small parameter. For primary resonance with
$\Omega \sim \omega_0$ in the linear regime, where both damping
and non-linearities are negligible, the amplitude growing is
unbounded according to $A \sim t$, but $\Omega = \omega_0$ exactly
(Landau and Lifshitz 1976). The linear growing of amplitude is
limited by damping and non-linearities and detuning of the
frequencies is allowed. We describe the frequency detuning by a
detuning parameter $\sigma$ due to
\begin{equation}
\Omega = \omega_0+\epsilon \sigma.
\end{equation}
The frequency-response equation, relating the amplitude of the
resulting oscillations $A$ and $\sigma$ in dependence on the
amplitude of the excitation $k$, and the parameters $\mu$ and
$\alpha$, can be put into the form (Nayfeh and Mook 1979)
\begin{equation}\label{frkv-resp-eq1}
\sigma=\frac{3\alpha}{8\omega_0}\,A^2\pm
\left(\frac{k^2}{4\omega_0^2\,A^2}-\mu^2\right)^{1/2}.
\end{equation}
The linear response ($\alpha=0$) is symmetric, while $\alpha > 0$
introduces an asymmetry -- in both cases the peak amplitude
\begin{equation}
A_{\mathrm{max}}=\frac{k}{2\omega_0\,\mu};
\end{equation}
(see Nayfeh and Mook 1979 for details). The maximal frequency
detuning allowed for resonant phenomena represents few percent of
the eigenfrequency $\omega_0$.

In the case of superharmonic resonances
\begin{equation}
3\Omega=\omega_0+\epsilon \sigma,
\end{equation}
the frequency-response equation reads
\begin{equation}\label{frkv-resp-eq2}
\sigma=3\frac{\alpha\,
\Lambda^2}{\omega_0}+\frac{3\alpha}{8\omega_0}\,A^2\pm
\left(\frac{\alpha^2\Lambda^6}{\omega_0^2\,A^2}-\mu^2\right)^{1/2},
\end{equation}
where
\begin{equation}
\Lambda=\frac{k}{2\left(\omega_0^2-\Omega^2\right)}.
\end{equation}
The peak amplitude of the resonant overtone oscillation is
\begin{equation}
A_{\mathrm{max}}=\frac{\alpha\,\Lambda^3}{\omega_0\,\mu},
\end{equation}
\textit{i.e.}, it depends on the magnitude of the non-linear term,
contrary to the case of primary resonances. Similar
frequency-response equations can be given for the subharmonic
resonance ($\Omega=3\omega_0+\epsilon\sigma$) or for combinational
resonances (Nayfeh and Mook 1979). Generally, the maximal
frequency scatter represents few percent of $\omega_0$ again
(Nayfeh and Mook 1979). Similarly, in the case of the quadratic
non-linearity ($-\epsilon\,\alpha\,u^2$) introducing the
sub(super)harmonic frequencies $\Omega\sim 2\omega_0$
($\Omega\sim\frac{1}{2}\omega_0$).

In the case of the parametric resonance, where the excitation
appears as a time-dependent coefficient in oscillatory equations,
and for internal resonances between coupled oscillatory modes, the
frequency detuning is given in a more complex way (Nayfeh and Mook
1979, Landau and Lifshitz 1976) but the maximal relative frequency
detuning is again in percents and sharply decreases with the order
of the resonance (Landau and Lifshitz 1976). We shall not go into
details leaving them to future work. Here we concentrate on the
resonant conditions.

The formulae for the vertical epicyclic frequency $\nu_{\theta}$
and the radial epicyclic frequency $\nu_{\rm r}$ take in the
gravitational field of a rotating Kerr black hole (with the mass
$M$ and dimensionless spin $a$) the form (\eg Aliev and Galtsov
1981, Kato \etal 1998)
\begin{equation}
\label{frequencies} \nu_{\theta}^2 =
\alpha_\theta^{\phantom{1}}\,\nu_\mathrm{K}^2, \qquad \nu_{\rm
r}^2 = \alpha_\mathrm{r}^{\phantom{1}}\nu_\mathrm{K}^2
\end{equation}
where the Keplerian frequency and related dimensionless epicyclic
frequencies are given by the formulae
\begin{eqnarray}
\nu_{\mathrm{K}}&=&\frac{1}{2\pi}\left(\frac{\mathrm{G}M}{r_\mathrm{G}^{~3}}\right)^{1/2}
\left(x^{3/2} + a \right)^{-1} =
\frac{1}{2\pi}\left(\frac{\mathrm{c}^3}{\mathrm{G}M}\right)
\left(x^{3/2}+a\right)^{-1},\nonumber\\
\alpha_\theta&=& 1-4\,a\,x^{-3/2}+3a^2\,x^{-2},
\nonumber\\
\alpha_\mathrm{r}&=&1-6\,x^{-1}+ 8 \,a \, x^{-3/2} -3 \, a^2 \,
x^{-2}.
\end{eqnarray}
Here $x = r/(\mathrm{G}M/\mathrm{c}^2)$ is the dimensionless
radius, expressed in terms of the gravitational radius of the
black hole.

For a particular resonance $n:m$ the equation
\begin{equation}
\label{ratios} n\nu_{\rm r} = m\nu_{\rm v};
\quad\nu_\mathrm{v}\in\{\nu_\theta,\, \nu_\mathrm{K}\}
\end{equation}
determines the dimensionless resonance radius $x_{n:m}$ as a
function of the dimensionless spin $a$ in the case of direct
resonances that can be easily extended to the resonances with
combinational frequencies. From the known mass of the central
black hole (\eg low-mass in the case of binary systems or
high-mass in the case of supermassive black holes), and the
observed twin peak frequencies ($\nu_{\rm upp}$, $\nu_{\rm
down}$), the Eqs (\ref{frequencies})\,--\,(\ref{ratios}) imply the
black hole spin $a$ for various versions of the resonance model,
with the beat frequencies taken into account (T{\"{o}}r{\"{o}}k
\etal 2005).

\Section{Black Holes Allowing Multiple Resonances}
Generally, the
resonances could be excited at different radii of the accretion
disc under different internal conditions; such a situation is
discussed in detail by Stuchl{\'{\i}}k \etal (2007b). On the other
hand, physically more interesting case arises when two (or more)
resonances are excited at a common radius because of probable
internal physical connection of those resonances. Of course, a
causal connection of two resonances could hardly be relevant, but
is not excluded, if they appear at distant radii.

A simple possibility of two resonances sharing a common radius can
be conceived supposing a direct resonance and some combinational
resonances of the frequencies entering the direct resonances. For
example, in the field of any Kerr black hole a direct resonance of
the epicyclic frequencies $\nu_{\theta}:\nu_{\mathrm{r}}=3:2$
admits combinational resonances
$\nu_{\theta}:(\nu_{\theta}-\nu_{\mathrm{r}})=3:1$ and
$\nu_{\mathrm{r}}:(\nu_{\theta}-\nu_{\mathrm{r}})=2:1$ (see, \eg
Stuchl{\'{\i}}k \etal 2007b). However, this kind of behaviour is
not exhibited by any of the four microquasars investigated at
present time (T{\"{o}}r{\"{o}}k \etal 2005).

Nevertheless, it is important that there exists a possibility of
direct resonances of oscillations with all of the three orbital
frequencies, characterized by a triple frequency ratio set
\begin{equation}\label{pomer}
             \nu_{\mathrm{K}}:\nu_{\theta}:\nu_{\mathrm{r}} = s:t:u
\end{equation}
with $s>t>u$ being small integers. The frequency set ratio (Eq.
\ref{pomer}) can be realized only for special values of the black
hole spin $a$. The black hole mass is then related to the
magnitude of the frequencies.

In order to look for the special values of the black hole spin $a$
and the related frequency ratios (Eq.~\ref{pomer}), we have to
summarize properties of the Keplerian and epicyclic frequencies of
the circular geodesic motion in the Kerr spacetimes. It is well
known that $\nu_{\mathrm{K}}$ and $\nu_{\theta}$ are defined down
to the photon circular geodesic located at $x_{\mathrm{ph}}$ that
is determined by the condition
\begin{equation}
             a = a_{\mathrm{ph}}(x)\equiv \frac{\sqrt{x}}{2}(3-x)
\end{equation}
while $\nu_{\mathrm{r}}$ is defined down to the innermost
(co-rotating) stable circular geodesic at $x_{\mathrm{ms}}$ given
by
\begin{equation}
             a = a_{\mathrm{ms}}(x)\equiv \frac{\sqrt{x}}{3}\left(4-\sqrt{3x-2}\right).
\end{equation}
Clearly, we have to search for the integer ratios (Eq.
\ref{pomer}) at $x \geq x_{\mathrm{ms}}$.

The Keplerian frequency $\nu_{\mathrm{K}}(r,a)$ is a monotonically
decreasing function of the radial coordinate for any value of the
black hole spin. On the other hand, the radial epicyclic frequency
has the global maximum for any Kerr black hole. The vertical
epicyclic frequency is not monotonic, if the spin is sufficiently
high (see, \eg Kato \etal 1998, Perez \etal 1997). For the Kerr
black hole spacetimes, the locations
$\mathcal{R}\,_\mathrm{r}(a),~\mathcal{R}\,_\theta(a)$ of maxima
of the epicyclic frequencies $\nu_\mathrm{r},~\nu_\theta$ are
implicitly given by the conditions (T{\"{o}}r{\"{o}}k and
Stuchl{\'{\i}}k 2005a)
\begin{eqnarray}
\label{implicitcondition}
\beta_{j}(x,a)&=&\frac{1}{2}\frac{\sqrt{x}}{x^{3/2}+a}\alpha_{j}(x,a)\quad \mathrm{where}~j\in\{\mathrm{r},\theta\},\nonumber\\
\beta_{\mathrm{r}}(x,a)&\equiv&\frac{1}{x^{2}}-\frac
{2a}{x^{5/2}}+ \frac {a^2}
{x^3},\nonumber\\
\beta_{\theta}(x,a)&\equiv&\frac {a}{x^{5/2}}-\frac {a^2}{x^3}.
\end{eqnarray}

For any black hole spin, the extrema of the radial epicyclic
frequency $\mathcal{R}\,_\mathrm{r}(a)$ must be located above the
marginally stable orbit. On the other hand, the latitudinal
extrema $\mathcal{R}\,_\theta(a)$ are located above the photon
(marginally bound or marginally stable) circular orbit only if the
limits on the black hole spin $a>0.748$ (0.852, 0.952) are
satisfied (T{\"{o}}r{\"{o}}k and Stuchl{\'{\i}}k 2005b). In the
Keplerian discs, with the inner boundary $x_{\mathrm{in}} \sim
x_{\mathrm{ms}}$, the limiting value $a=0.952$ is relevant.

From the point of view of the observational consequences, it is
important to know, for which frequency ratios $n$\,:\,$m$ the
resonant frequencies $\nu_{\theta}(a,n\!:\! m)$, considered as a
function of the black hole spin $a$ for a given frequency ratio
$n$\,:\,$m$, has a non-monotonic character. A detailed analysis
(T{\"{o}}r{\"{o}}k and Stuchl{\'{\i}}k 2005a) shows that
$\nu_{\theta}(a,n\!:\! m)$ has a local maximum for $n$\,:\,$m
>$ 11\,:\,5, \ie in physically relevant situations ($n$, $m$
small enough for the resonance), it occurs for the ratios
$\nu_{\theta}:\nu_{\mathrm{r}}=3\!:\!1$, $4\!:\!1$, $5\!:\!2$,
$5\!:\!1$.

Assuming two resonances $\nu_{\mathrm{K}}:\nu_{\theta}=s:t$ and
$\nu_{\mathrm{K}}:\nu_{\mathrm{r}}=s:u$ occurring at the same $x$,
we arrive to the conditions
\begin{equation}
    \alpha_{\theta}(a,x)=\left(\frac{t}{s}\right)^2,
\end{equation}
\begin{equation}
    \alpha_{\mathrm{r}}(a,x)=\left(\frac{u}{s}\right)^2
\end{equation}
that have to be solved simultaneously for $x$ and $a$. The
solution is given by the condition
\begin{equation}
    a^{\theta}(x,t/s)=a^{\mathrm{r}}(x,u/s)
\end{equation}
where
\begin{equation}\label{spin}
    a^{\theta}(x,t/s)\equiv \frac{\sqrt{x}}{3}\left(2\pm\sqrt{4-3x\left[1-\left(\frac{t}{s}\right)^2\right]}\right),
\end{equation}
\begin{equation}
    a^{\mathrm{r}}(x,u/s)\equiv \frac{\sqrt{x}}{3}\left(4\pm\sqrt{-2+3x\left[1-\left(\frac{u}{s}\right)^2\right]}\right).
\end{equation}
It is possible to find an explicit solution determining the
relevant radius for any triple frequency set ratio $s:t:u$
\begin{equation}
x(s,t,u)=\frac{6 s^2}{6 s^2\pm2 \sqrt{2} \sqrt{(t-u) (t+u) \left(3
s^2-t^2-2 u^2\right)}- \left(t^2+5 u^2\right)}.
\end{equation}

Clearly, the condition $t^2+2u^2 \leq 3s^2$ is always satisfied.
The corresponding black hole spin $a$ is then determined, \eg by
Eq.~(\ref{spin}) giving $a^{\theta}(x(s,t,u),t/s)$. Of course, we
consider only the black hole cases when $a \leq 1$. This condition
puts a restriction on allowed values of $s,t,u$.

The solutions have been found for frequency ratios with $s \leq
5$. We were looking for the black hole spin $a$ allowing ratios
$s:t:u = 3\!:\!2\!:\!1$, $4\!:\!3\!:\!2$, $4\!:\!3\!:\!1$,
$4\!:\!2\!:\!1$, $5\!:\!4\!:\!3$, $5\!:\!4\!:\!2$,
$5\!:\!4\!:\!1$, $5\!:\!3\!:\!2$, $5\!:\!3\!:\!1$,
$5\!:\!2\!:\!1$. Higher values of the $s,t,u$ are presented and
discussed in a more general context of triple frequency set ratios
that could arise when two resonances occur at two different radii,
but for special values of the black hole spin two of the frequency
levels (top, bottom, mixed) coincide (see Stuchl{\'{\i}}k \etal
2007b).

\begin{figure}[t]
\begin{center}
\subfigure[][]{\includegraphics[width=.48\hsize]{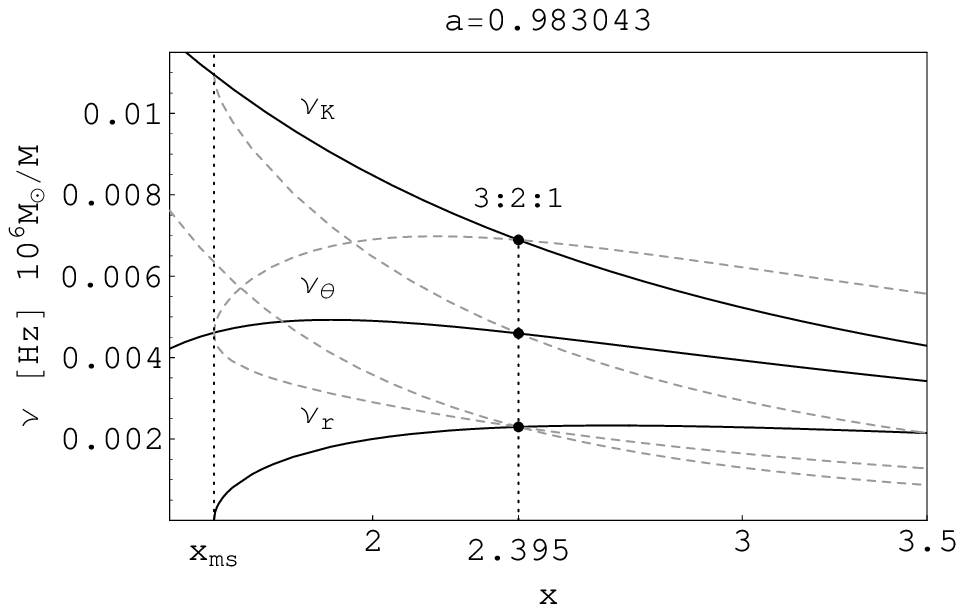}\label{triplety-a}}\quad
\subfigure[][]{\includegraphics[width=.48\hsize]{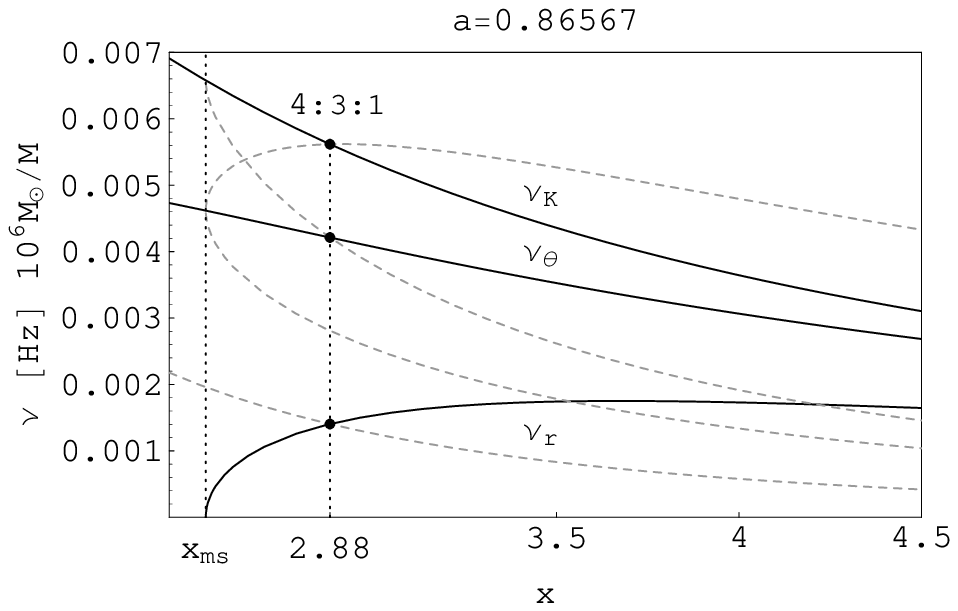}\label{triplety-b}}\\
\subfigure[][]{\includegraphics[width=.48\hsize]{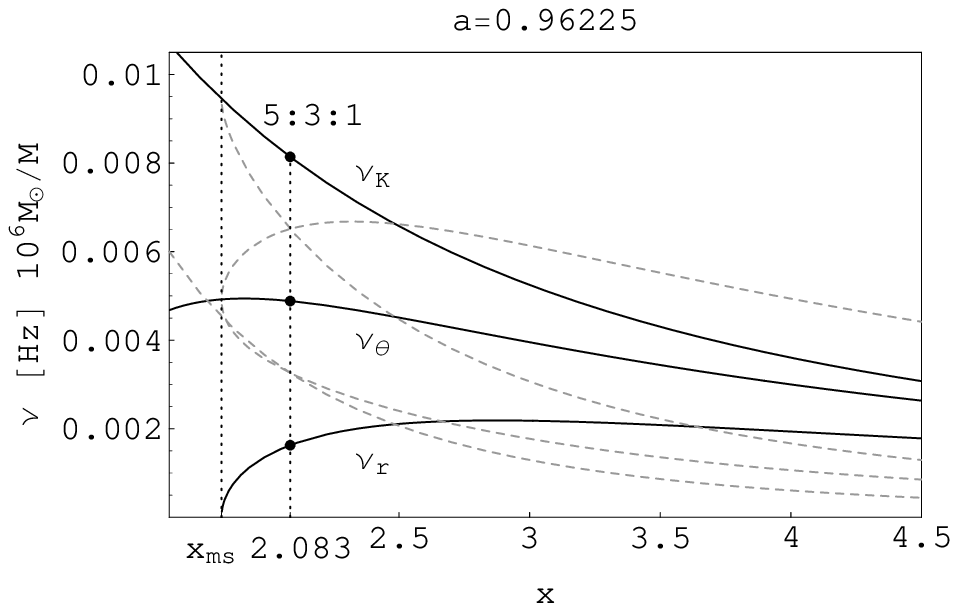}\label{triplety-c}}\quad
\subfigure[][]{\includegraphics[width=.48\hsize]{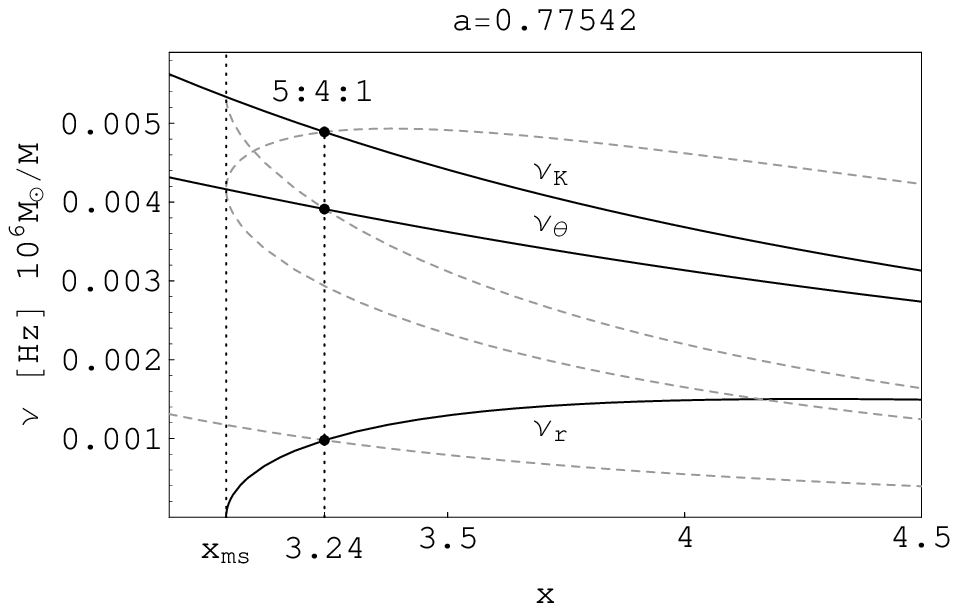}\label{triplety-d}}\\
\subfigure[][]{\includegraphics[width=.48\hsize]{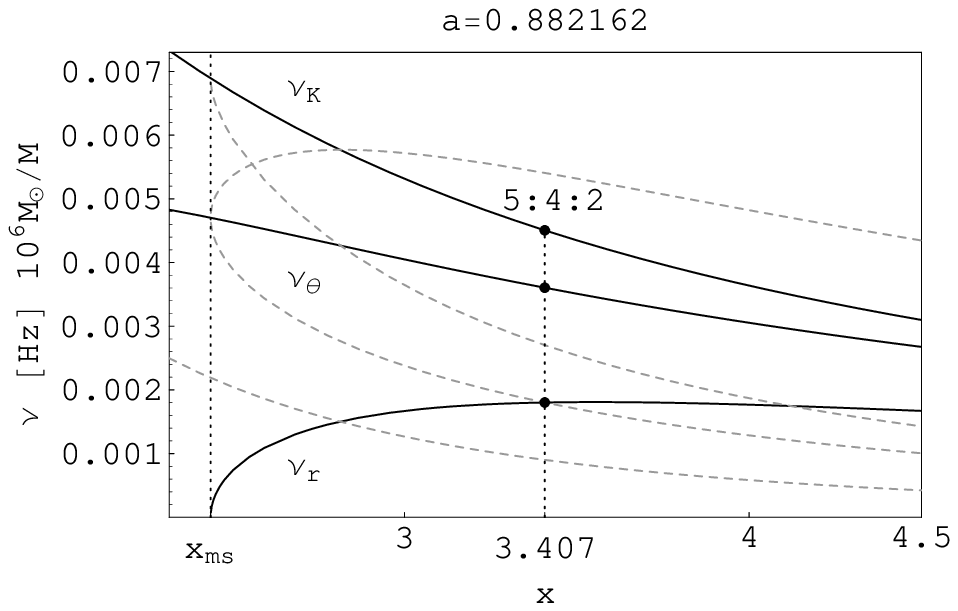}\label{triplety-e}}\quad
\subfigure[][]{\includegraphics[width=.48\hsize]{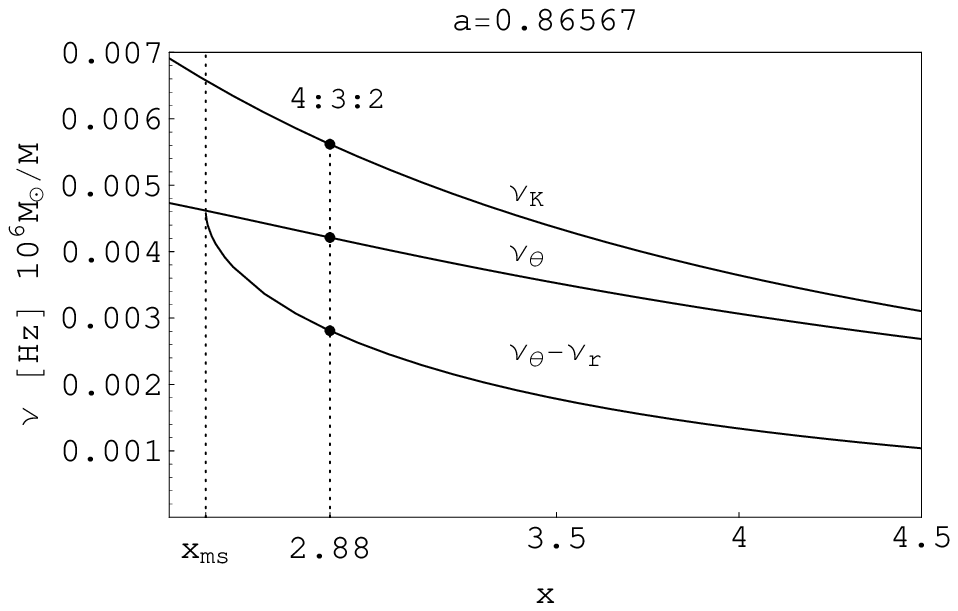}\label{triplety-f}}
\end{center}
\FigCap{Black holes enabling interrelated strong resonances. The
special cases of the triple frequency ratio sets of the orbital
frequencies $\nu_{\mathrm{K}}$, $\nu_{\theta}$, $\nu_{\mathrm{r}}$
(black solid lines) with the corresponding spin $a$ and the shared
resonance radius. For completeness we present the relevant simple
combinational frequencies $\nu_{\theta}-\nu_{\mathrm{r}}$,
$\nu_{\theta}+\nu_{\mathrm{r}}$, $\nu_{\mathrm{K}}-\nu_{\theta}$,
$\nu_{\mathrm{K}}-\nu_{\mathrm{r}}$ (grey dashed lines). Notice
that the extraordinary resonant spin $a_{3:2:1}=0.983$ represents
the only case when the combinational and direct orbital
frequencies coincide at the shared resonance
radius.\label{triplety}}
\end{figure}

We have shown that the direct resonances could result only in the
triple frequency sets $3\!:\!2\!:\!1$, $4\!:\!3\!:\!1$,
$5\!:\!4\!:\!2$, $5\!:\!4\!:\!1$, $5\!:\!3\!:\!1$ with values of
the spin and the radius $x_{s:t:u}$ given in
Fig.~\ref{triplety}a\,--\,e. The other ratios can be realized with
the combinational frequencies involved in the triple frequency
sets $4\!:\!3\!:\!2$, $4\!:\!2\!:\!1$, $5\!:\!4\!:\!3$,
$5\!:\!3\!:\!2$, $5\!:\!2\!:\!1$ as shown in Fig.~\ref{triplety}f
for the case of the set $4\!:\!3\!:\!2$. We can summarize the
results in the following way.

\begin{enumerate}
 \item[a)]
        $\nu_{\mathrm{K}}:\nu_{\theta}:\nu_{\mathrm{r}} = 3:2:1$ (see Fig.~\ref{triplety}a)
        \begin{equation}
        a_{3:2:1}=0.983,\qquad x_{3:2:1}=2.395,\qquad x_{\mathrm{ms}}=1.571.
        \end{equation}
    This case of the so called extraordinary resonant spin
    $a_{3:2:1}=0.983$ may allow strong resonances, in situations when
    the Keplerian and epicyclic frequencies are in the lowest possible
    ratio at the common radius $x_{3:2:1} = 2.395$ (see
    Fig.~\ref{triplety}a), and is thus of special interest. In fact, this case involves
    the lowest structure of direct resonances with
    $\nu_{\mathrm{K}}:\nu_{\mathrm{r}}=3:1$, $\nu_{\mathrm{K}}:\nu_{\theta}=3:2$, $\nu_{\theta}:\nu_{\mathrm{r}}=2:1$.
    Notice that in this special case
    also any of the simple combinational frequencies coincides with one of the frequencies
    $\nu_{\mathrm{K}}$, $\nu_{\theta}$, $\nu_{\mathrm{r}}$, and are in the fixed small
    integer ratios
            \begin{equation}
            \frac{\nu_\mathrm{K}}{\nu_{\theta}-\nu_{\mathrm{r}}}=\frac{3}{1},\qquad
            \frac{\nu_{\mathrm{K}}}{\nu_{\mathrm{K}}-\nu_\mathrm{r}}=\frac{3}{2},\qquad
            \frac{\nu_{\theta}}{\nu_{\theta}-\nu_\mathrm{r}}=\frac{2}{1}.
            \end{equation}
    It should be stressed that this is the only case when the
    combinational frequencies not exceeding the Keplerian frequency
    are in the same ratios as the orbital frequencies. We obtain
    the strongest possible resonances when the beat
    frequencies enter the resonances satisfying the conditions
            \begin{equation}
            \frac{\nu_{\theta}+\nu_{\mathrm{r}}}{\nu_\mathrm{K}}=\frac{3}{3}=1,\quad
            \frac{\nu_{\theta}}{\nu_\mathrm{K}-\nu_{\mathrm{r}}}=\frac{2}{2}=1,\quad
            \frac{\nu_{\mathrm{r}}}{\nu_\mathrm{K}-\nu_{\theta}}=1,\quad
            \frac{\nu_{\theta}-\nu_\mathrm{r}}{\nu_{\mathrm{r}}}=1.
            \end{equation}
\item[b)]
    $\nu_{\mathrm{K}}:\nu_{\theta}:\nu_{\mathrm{r}} = 4:3:1$ (see Fig.~\ref{triplety}b)
        \begin{equation}
        a_{4:3:1}=0.866,\qquad x_{4:3:1}=2.88,\qquad x_{\mathrm{ms}}=2.539.
        \end{equation}
    In this case the combinational frequencies give additional frequency
    ratios
            \begin{equation}
        \frac{\nu_\mathrm{K}}{\nu_\mathrm{K}-\nu_{\mathrm{r}}}=\frac{4}{3},\quad
        \frac{\nu_{\theta}}{\nu_{\theta}-\nu_{\mathrm{r}}}=\frac{3}{2},\quad
        \frac{\nu_\mathrm{K}}{\nu_{\theta}-\nu_{\mathrm{r}}}=\frac{4}{2}=\frac{2}{1},\quad
        \frac{\nu_{\theta}-\nu_{\mathrm{r}}}{\nu_{\mathrm{r}}}=\frac{2}{1},
            \end{equation}
    and the strongest resonant ratios
        \begin{equation}
        \frac{\nu_{\theta}+\nu_{\mathrm{r}}}{\nu_\mathrm{K}}=\frac{4}{4}=1,\qquad
        \frac{\nu_{\theta}}{\nu_\mathrm{K}-\nu_{\mathrm{r}}}=\frac{3}{3}=1,\qquad
        \frac{\nu_{\mathrm{r}}}{\nu_\mathrm{K}-\nu_{\theta}}=1.
        \end{equation}
        Here, using the combinational frequency $\nu_{\theta}-\nu_{\mathrm{r}}$, we obtain
        the other three frequency ratio sets (see Fig.~\ref{triplety}f)
        \begin{equation}
        \nu_{\mathrm{K}}:\nu_{\theta}:\left(\nu_{\theta}-\nu_{\mathrm{r}}\right)=4:3:2
        \end{equation}
        and
        \begin{equation}
        \nu_{\mathrm{K}}:\left(\nu_{\theta}-\nu_{\mathrm{r}}\right):\nu_{\mathrm{r}}=4:2:1.
    \end{equation}
    Now, the four observable frequency ratio set is possible when the combinational
    frequency is mixed with the orbital frequencies
        \begin{equation}
    \left(\nu_\mathrm{K} = \nu_{\theta}+\nu_{\mathrm{r}}\right):
    \left(\nu_{\theta}=\nu_\mathrm{K}-\nu_{\mathrm{r}}\right):
    \left(\nu_{\theta}-\nu_{\mathrm{r}}\right):\left(\nu_{\mathrm{r}}
    = \nu_\mathrm{K}-\nu_{\theta}\right)=4:3:2:1.
        \end{equation}
\item[c)]
    $\nu_{\mathrm{K}}:\nu_{\theta}:\nu_{\mathrm{r}} = 5:3:1$ (see Fig.~\ref{triplety}c)
        \begin{equation}
        a_{5:3:1}=0.962,\qquad x_{5:3:1}=2.083,\qquad
        x_{\mathrm{ms}}=1.820,
        \end{equation}
    and the combinational frequencies (not exceeding
    $\nu_{\mathrm{K}}$) are in the ratios
\begin{eqnarray}
        &&\frac{\nu_\mathrm{K}}{\nu_\mathrm{K}-\nu_{\mathrm{r}}}=\frac{\nu_\mathrm{K}}{\nu_{\theta}+\nu_{\mathrm{r}}}=\frac{5}{4},\qquad
        \frac{\nu_\mathrm{K}-\nu_{\mathrm{r}}}{\nu_{\theta}}=\frac{4}{3},\qquad
        \frac{\nu_\theta}{\nu_{\theta}-\nu_{\mathrm{r}}}=\frac{3}{2},\nonumber\\
        &&\frac{\nu_\mathrm{K}-\nu_{\theta}}{\nu_{\mathrm{r}}}=\frac{\nu_\theta-\nu_{\mathrm{r}}}{\nu_{\mathrm{r}}}=\frac{2}{1},\qquad
        \frac{\nu_\mathrm{K}}{\nu_{\theta}-\nu_{\mathrm{r}}}=\frac{5}{2}.
        \end{eqnarray}
    Then we can generate triple frequency sets involving the
    combinational frequencies with
        the ratios in the form
        \begin{equation}
        \nu_{\mathrm{K}}:\left(\nu_{\theta}-\nu_{\mathrm{r}}\right):\nu_{\mathrm{r}}=5:2:1
        \end{equation}
    and
         \begin{equation}
            \nu_{\mathrm{K}}:\nu_{\theta}:\left(\nu_{\theta}-\nu_{\mathrm{r}}\right)=5:3:2.
        \end{equation}
    Moreover, using the combinational frequencies we could obtain two sets of four frequency
    ratios
        \begin{eqnarray}
   && \nu_\mathrm{K} :
    \left(\nu_\mathrm{K}-\nu_{\mathrm{r}}=\nu_{\theta}+\nu_{\mathrm{r}}\right)
    : \nu_{\theta} : \nu_{\mathrm{r}}=5:4:3:1,\nonumber\\
   && \nu_\mathrm{K} : \nu_{\theta} :
    \left(\nu_{\theta}-\nu_{\mathrm{r}}
    =\nu_\mathrm{K}-\nu_{\theta}\right) :
    \nu_{\mathrm{r}}=5:3:2:1,
        \end{eqnarray}
    and one set of five frequency ratio
         \begin{equation}
    \nu_\mathrm{K} :
    \left(\nu_\mathrm{K}-\nu_{\mathrm{r}}=\nu_{\theta}+\nu_{\mathrm{r}}\right)
    : \nu_{\theta} : \left(\nu_{\theta}-\nu_{\mathrm{r}} =
    \nu_\mathrm{K}-\nu_{\theta}\right) : \nu_{\mathrm{r}}=5:4:3:2:1.
    \end{equation}
\item[d)]
    $\nu_{\mathrm{K}}:\nu_{\theta}:\nu_{\mathrm{r}} = 5:4:1$ (see Fig.~\ref{triplety}d)
        \begin{equation}
        a_{5:4:1}=0.775,\qquad x_{5:4:1}=3.240,\qquad
        x_{\mathrm{ms}}=3.033.
        \end{equation}
    The combinational frequencies can give the ratios
         \begin{equation}
        \frac{\nu_\mathrm{K}}{\nu_\mathrm{K}-\nu_{\mathrm{r}}}=\frac{5}{4},\quad
        \frac{\nu_\theta}{\nu_\theta-\nu_{\mathrm{r}}}=\frac{4}{3},\quad
        \frac{\nu_\mathrm{K}}{\nu_{\theta}-\nu_{\mathrm{r}}}=\frac{5}{3},\quad
        \frac{\nu_{\theta}-\nu_{\mathrm{r}}}{\nu_{\mathrm{r}}}=\frac{3}{1},
    \end{equation}
    and the strongest resonant ratios
         \begin{equation}
        \frac{\nu_{\theta}+\nu_{\mathrm{r}}}{\nu_\mathrm{K}}=\frac{5}{5}=1,\qquad
        \frac{\nu_{\theta}}{\nu_\mathrm{K}-\nu_{\mathrm{r}}}=\frac{4}{4}=1,\qquad
        \frac{\nu_{\mathrm{r}}}{\nu_\mathrm{K}-\nu_{\theta}}=1.
        \end{equation}
        This case leads to the triple frequency ratio sets
\begin{eqnarray}
    &&\nu_{\mathrm{K}}:\left(\nu_{\theta}-\nu_{\mathrm{r}}\right):\nu_{\mathrm{r}}
    = 5:3:1,\nonumber\\
   && \nu_{\mathrm{K}}:\nu_{\theta}:\left(\nu_{\theta}-\nu_{\mathrm{r}}\right)=5:4:3.
        \end{eqnarray}
    Here, only one of the four frequency ratio sets is possible, namely
        \begin{equation}
    \left(\nu_\mathrm{K} = \nu_{\theta} +
    \nu_{\mathrm{r}}\right) :
    \left(\nu_{\theta}=\nu_\mathrm{K}-\nu_{\mathrm{r}}\right)
    :\left(\nu_{\theta}-\nu_\mathrm{r}\right) : \nu_{\mathrm{r}}=5:4:3:1.
        \end{equation}
\item[e)]
    $\nu_{\mathrm{K}}:\nu_{\theta}:\nu_{\mathrm{r}} = 5:4:2$ (see Fig.~\ref{triplety}e)
        \begin{equation}
        a_{5:4:2}=0.882,\qquad x_{5:4:2}=3.407,\qquad
        x_{\mathrm{ms}}=2.438.
        \end{equation}
    Here, the combinational frequencies give the ratios
\begin{eqnarray}
        &&\frac{\nu_{\theta}}{\nu_\mathrm{K}-\nu_{\mathrm{r}}}=\frac{4}{3},\qquad
        \frac{\nu_\mathrm{K}}{\nu_\mathrm{K}-\nu_{\mathrm{r}}}=\frac{5}{3},\qquad
        \frac{\nu_{\theta}}{\nu_\theta-\nu_{\mathrm{r}}}=\frac{4}{2}=\frac{2}{1},\nonumber\\
        &&\frac{\nu_{\mathrm{r}}}{\nu_\mathrm{K}-\nu_{\theta}}=\frac{2}{1},\qquad
        \frac{\nu_\mathrm{K}}{\nu_{\theta}-\nu_{\mathrm{r}}}=\frac{5}{2},
\end{eqnarray}
    and the strongest resonant ratio
        \begin{equation}
        \frac{\nu_{\theta}-\nu_{\mathrm{r}}}{\nu_{\mathrm{r}}}=\frac{2}{2}=1.
        \end{equation}
        This case leads to the triple frequency ratio sets
    \begin{eqnarray}
        &&\nu_{\theta}:\left(\nu_{\mathrm{K}}-\nu_{\mathrm{r}}\right):\nu_{\mathrm{r}}=4:3:2,\nonumber\\
        &&\nu_{\mathrm{K}}:\nu_{\theta}:\left(\nu_{\mathrm{K}}-\nu_{\mathrm{r}}\right)=5:4:3,
    \end{eqnarray}
        and again the related four frequency ratio sets
    \begin{eqnarray}
    &&\nu_\mathrm{K} : \nu_{\theta}:
    \left(\nu_\mathrm{K}-\nu_{\mathrm{r}}\right):
    \left(\nu_{\mathrm{r}}=\nu_{\theta}-\nu_{\mathrm{r}}\right)=5:4:3:2,\nonumber\\
   && \nu_\mathrm{K} : \nu_{\theta}:
    \left(\nu_{\mathrm{r}}=\nu_{\theta}-\nu_{\mathrm{r}}\right) :
    \left(\nu_\mathrm{K}-\nu_{\theta}\right)=5:4:2:1,
    \end{eqnarray}
    and one five frequency ratio set
        \begin{equation}
    \nu_\mathrm{K} : \nu_{\theta}:
    \left(\nu_\mathrm{K}-\nu_{\mathrm{r}}\right):
    \left(\nu_{\mathrm{r}}=\nu_{\theta}-\nu_{\mathrm{r}}\right) :
    \left(\nu_\mathrm{K}-\nu_{\theta}\right)=5:4:3:2:1.
        \end{equation}
\end{enumerate}

\begin{figure}[t]
\begin{center}
\includegraphics[width=.5\hsize]{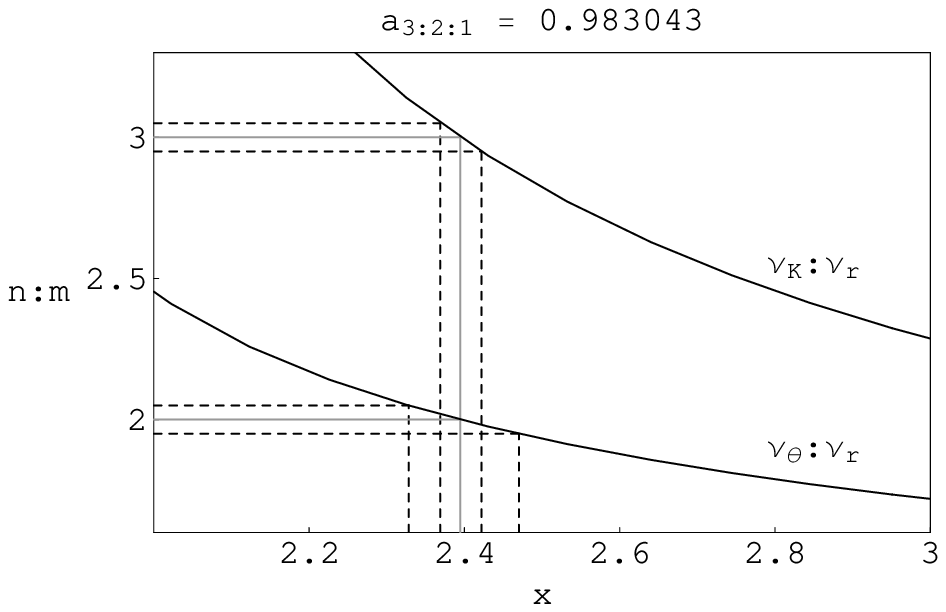}
\end{center}
\FigCap{\label{graf-rozptyl-pol}Calculation of the width of the
resonant radius where the strong resonant phenomena are possible.
Black lines represent the ratio $n\!:\!m =
\nu_{\mathrm{K}}\!:\!\nu_{\mathrm{r}}$,
$\nu_{\theta}\!:\!\nu_{\mathrm{r}}$, dashed lines illustrate the
relative frequency detuning $\delta = 5\%$ (\textit{i.e.},
$n\!:\!m = 2\pm 0.05$ and $3\pm 0.05$). Gray solid lines represent
the extraordinary case where the frequency ratio
$\nu_{\mathrm{K}}\!:\!\nu_{\theta}\!:\!\nu_{\mathrm{r}}=3:2:1$
arises at the same radius $x_{3:2:1}=2.395$. We can simply
calculate that at $x=2.369$ the frequency ratio is
$\nu_{\mathrm{K}}\!:\!\nu_{\theta}\!:\!\nu_{\mathrm{r}}=3.05:2.019:1$
and at $x=2.422$, there is
$\nu_{\mathrm{K}}\!:\!\nu_{\theta}\!:\!\nu_{\mathrm{r}}=2.95:1.981:1$.} 
\end{figure}

\begin{figure}[t]
\begin{center}
\subfigure{\includegraphics[width=.48\hsize]{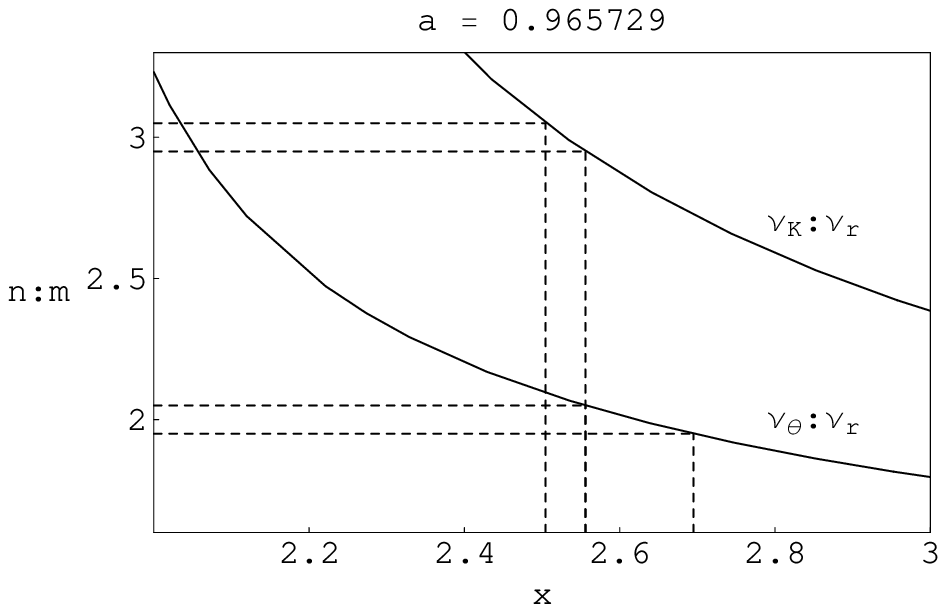}}\quad
\subfigure{\includegraphics[width=.48\hsize]{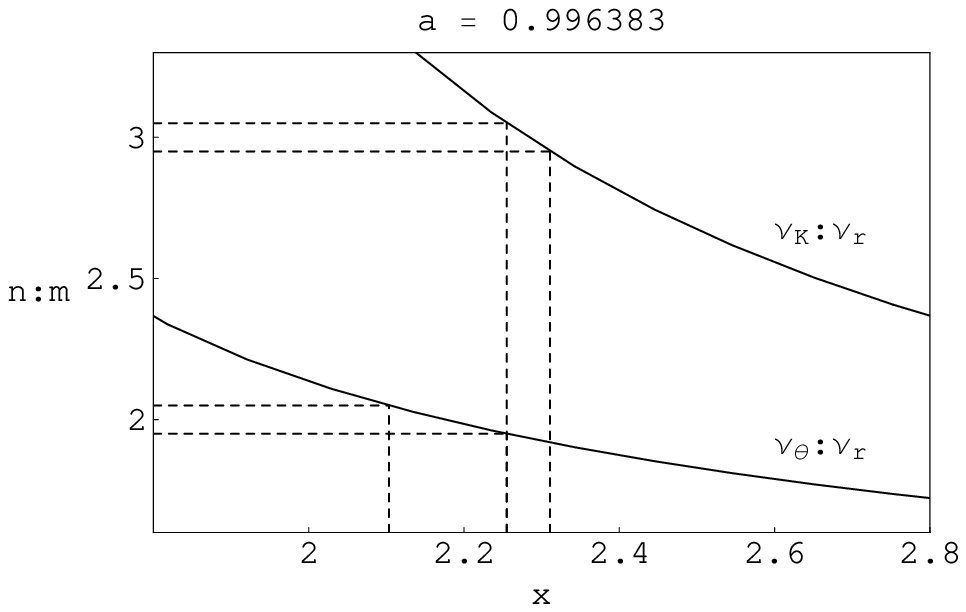}}
\end{center}
\FigCap{\label{graf-rozptyl}Calculation of the width of the
special value of the black hole spin parameter $a_{3:2:1}$ for
which the multiple values frequencies should be observed. Black
lines represent the ratio $n\!:\!m =
\nu_{\mathrm{K}}\!:\!\nu_{\mathrm{r}}$,
$\nu_{\theta}\!:\!\nu_{\mathrm{r}}$, dashed lines illustrate the
relative frequency detuning $\delta = 5\%$ (\textit{i.e.},
$n\!:\!m = 2\pm 0.05$ and $3\pm 0.05$). \textit{Left panel}: The
minimal value of the black hole spin parameter allowing the strong
resonant phenomena where
$\nu_{\mathrm{K}}\!:\!\nu_{\theta}\!:\!\nu_{\mathrm{r}}=2.95:2.05:1$
at $x=2.556$. \textit{Right panel}: The maximal value of the black
hole spin allowing the strong resonant phenomena where
$\nu_{\mathrm{K}}\!:\!\nu_{\theta}\!:\!\nu_{\mathrm{r}}=3.05:1.95:1$
at $x=2.255$.}
\end{figure}

Considering the extraordinary resonant spin $a=a_{3:2:1}=0.983$,
we can conclude that the resonant phenomena are possible in some
region around the resonant radius $x_{3:2:1}=2.395$ due to the
allowed frequency detuning with detuning parameter $\sigma$ given
by the frequency-response equations (see, \textit{e.g.},
Eq.~\ref{frkv-resp-eq1}, Eq.~\ref{frkv-resp-eq2}). The magnitude
of the frequency scatter $\Delta = \epsilon \sigma$ is fully
determined by a concrete resonance under conside\-ration and by
physical conditions in the oscillatory system, nevertheless, we
can put a general restriction on the allowed frequency detuning as
few percent of the oscillator eigenfrequency (Nayfeh and Mook
1979) and use it to find region of the disc around $x_{3:2:1}$,
where the resonance could appear. We introduce the relations
\begin{equation}
    \nu_1=2\nu_0\pm\Delta, \qquad \qquad \nu_2=3\nu_0\pm\Delta,
\end{equation}
\begin{equation}\label{podm-rozpt1}
    \frac{t}{u}=\frac{\nu_1}{\nu_0}=2\pm\delta,
    \end{equation}
\begin{equation}\label{podm-rozpt2}
    \frac{s}{u}=\frac{\nu_2}{\nu_0}=3\pm\delta,
    \end{equation}
where
\begin{equation}\label{def-delty}
    \delta\equiv\frac{\Delta}{\nu_0};
\end{equation}
the resonance could appear where the regions given by conditions
(\ref{podm-rozpt1}), (\ref{podm-rozpt2}) overlap (see
Fig.~\ref{graf-rozptyl-pol}). We find that the maximal frequency
detuning of $0.05$ implies possibility of the strong resonant
phenomena in the interval of $x\in \langle2.369;2.422\rangle$
extended around the radius $x_{3:2:1}=2.395$. Using this approach,
we can give, for a given relative frequency detuning
$\delta=\Delta/\nu_0$, the range of the black hole spin allowing
the strong resonant phenomena. We simply find the spin where the
overlap of the radial regions allowing resonance disappears (see
Fig.~\ref{graf-rozptyl}). We choose characteristic values of
$\delta = 0.01, 0.02$ and $0.05$ and using a simple numerical code
we determine the scatter around the special resonant spin values
$a_{3:2:1}$, $a_{4:3:1}$ and $a_{5:3:1}$. The results are
summarized in Table \ref{rozp-tab}. For the other special resonant
spins and any appropriately chosen value of $\delta$, the spin
scatter can be given by the numerical code. (We give the spin with
precision of $0.001$ as the $1\%$ frequency scatter implies the
spin scatter $\approx 0.001$.)

\renewcommand{\arraystretch}{1.6}
\MakeTable{|c|c|c|c|}{11cm}{\label{rozp-tab}Estimates of the spin
value range where the strong resonant phenomena are possible for
characteristic values of relative frequency detuning $\delta$.}
{\hline $\delta\ \left[\%\right]$ & $s:t:u = 3:2:1$ & $s:t:u = 4:3:1$ & $s:t:u = 5:3:1$ \\
\hline
$1$ & $a_{3:2:1}=0.983^{+0.003}_{-0.003}$ & $a_{4:3:1}=0.866^{+0.005}_{-0.005}$  & $a_{5:3:1}=0.962^{+0.001}_{-0.001}$ \\
$2$ & $a_{3:2:1}=0.983^{+0.006}_{-0.006}$ & $a_{4:3:1}=0.866^{+0.010}_{-0.010}$  & $a_{5:3:1}=0.962^{+0.003}_{-0.003}$ \\
$5$ & $a_{3:2:1}=0.983^{+0.013}_{-0.017}$ & $a_{4:3:1}=0.866^{+0.023}_{-0.027}$  & $a_{5:3:1}=0.962^{+0.006}_{-0.007}$ \\
 \hline}

\Section{A Possible Application to Sgr\,A$^*$} The resonant
phenomena could occur frequently, when different versions of the
re\-so\-nan\-ce model can be realized at a shared radius (or its
close vicinity) fixed by the frequency ratio
$\nu_{\mathrm{K}}:\nu_{\theta}:\nu_{\mathrm{r}} = s:t:u$.

As explored above, there is an extraordinary resonant black hole
spin $a_{3:2:1}=0.983$, when at the radius $x_{3:2:1}=2.395$ the
frequency ratio is $\nu_{\mathrm{K}}:\nu_{\theta}:\nu_{\mathrm{r}}
= 3:2:1$. Clearly, in vicinity of black holes with $a=0.983$, the
resonant phenomena should be strong, as the order of the
resonances is of the lowest possible values, and, moreover, all
the resonances, including those with beat frequencies, could
cooperate efficiently even for frequencies slightly scattered from
the exact resonant eigenfrequencies. The scatter of resonant
frequencies strongly depends on the order of resonances and can be
wide for resonances of very low order (Landau and Lifshitz 1976).

When the simple combinational frequencies could be considered, the
frequency ratios, \eg
$\nu_{\mathrm{r}}:\left(\nu_{\mathrm{K}}-\nu_{\theta}\right)=1:1$,
then allow strongest possible resonances.

\begin{figure}[t]
\begin{center}
\subfigure{\includegraphics[width=.48\hsize]{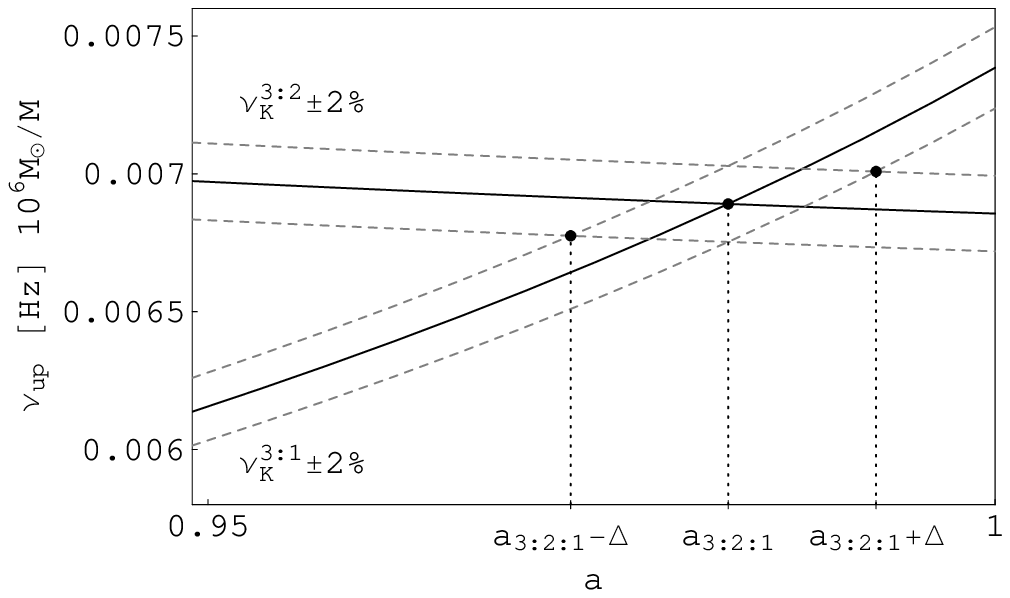}}\quad
\subfigure{\includegraphics[width=.48\hsize]{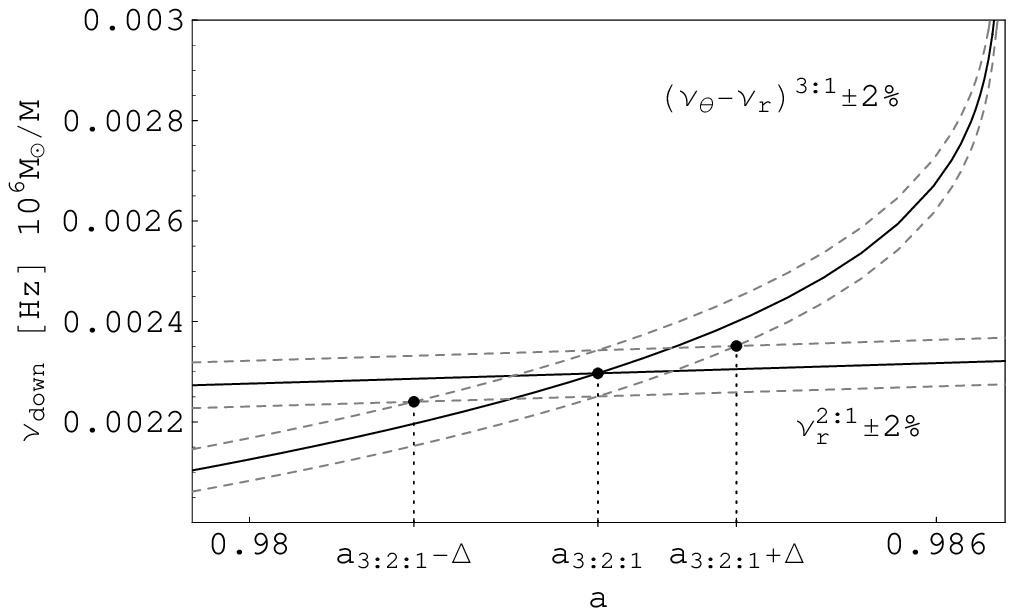}}
\end{center}
\FigCap{Error in determining the extraordinary resonant spin
$a_{3:2:1}=0.983$: the scatter of the black hole spin related to
the $2\%$ error in frequency measurements. It has to be confronted
with the spin scatter allowing occurrence of resonances (see Eqs
(\ref{podm-rozpt1})\,--\,(\ref{def-delty})). \textit{Left panel}:
The case of top identity $\nu_{\mathrm{upp}}=\nu_{\mathrm{K}}$:
the interval of allowed values of the spin is then $a\in \langle
0.973; 0.992 \rangle$. \textit{Right panel}: The case of bottom
identity
$\nu_{\mathrm{down}}=\nu_{\mathrm{r}}=\left(\nu_{\theta}-\nu_{\mathrm{r}}\right)$:
the interval of allowed values of the spin is $a\in \langle 0.981;
0.984\rangle$.\label{presnost}}
\end{figure}

That is the reason why we could expect well observable QPOs in the
field of black holes with spin close to the values allowing the
cooperating resonant phenomena sharing a fixed radius. It is
important to look for some candidate systems exploring the simple
triple frequency ratio sets. In such situations we can determine
the black hole spin with high precision, given by the frequency
measurement error (see Fig.~\ref{presnost}). This could help very
much in determining the other physical characteristics of the
systems. Notice that errors of frequency measurements imply some
errors in the spin determination, as illustrated in
Fig.~\ref{presnost}. It depends also on the concrete resonances
occurring at a given radius. Clearly, if the relevant frequency
curves cross in a large (small) relative angle, the spin is
determined with high (low) precision.

The Galaxy centre source Sgr\,A$^*$ can serve as a proper
candidate system, since three QPOs were reported (but not fully
accepted by the astrophysical community) for the system
(Aschenbach 2004, T{\"{o}}r{\"{o}}k 2005) with frequency ratio
corresponding to the extraordinary resonant spin
\begin{equation}
(1/692) : (1/1130) : (1/2178) \approx 3 : 2 : 1
\end{equation}
and with the upper frequency being observed with a rather high
error
\begin{equation}
      \nu_{\mathrm{upp}} = (1.445 \pm 0.16)~\mathrm{mHz}.
\end{equation}

Considering the standard epicyclic resonance model for the two
upper frequencies $\nu_{\theta}:\nu_{\mathrm{r}}=3:2$, and the
bottom frequency
$\nu_{\mathrm{down}}=\nu_{\theta}-\nu_{\mathrm{r}}$, we obtain
$\nu_{\theta}:\nu_{\mathrm{r}}:(\nu_{\theta}-\nu_{\mathrm{r}})=3:2:1$.
Then identifying $\nu_{\mathrm{upp}} = \nu_{\theta}$, we can show
that for $a \leq 1$, the black hole mass $M \leq 2.3 \times 10^6
~\mathrm{M}_{\odot}$, which is in clear disagreement with the
allowed range of the Sgr\,A$^*$ mass coming from the most recent
analysis of the orbits of stars moving within $0.01\,\mathrm{pc}$ 
of Sgr\,A$^*$ (Ghez \etal 2008)
\begin{equation}\label{mass-Ghez-1}
    3.5 \times 10^6 ~\mathrm{M}_{\odot} < M < 4.7 \times 10^6
    ~\mathrm{M}_{\odot}
\end{equation}
and with the estimate $M \approx 4 \times 10^6~\mathrm{M}_{\odot}$
obtained by Reid (2008).

If the black hole is assumed to be at rest with respect to the
Galaxy (\ie has no massive companion to induce motion), the fit
could be further constrained to the interval (Ghez \etal 2008)
\begin{equation}\label{mass-Ghez-2}
    4.1 \times 10^6 ~\mathrm{M}_{\odot} < M < 4.9 \times 10^6
    ~\mathrm{M}_{\odot}
\end{equation}
that is the most likely mass estimate at present, considering the
discovery that Sgr\,A$^*$ is nearly stationary at the Galactic
center (Reid 2008).

However, assuming a black hole with the spin comparable to the
extraordinary resonant value of $a \approx 0.983$, with the
frequency ratio $\nu_{\mathrm{K}}:\nu_{\theta}:\nu_{\mathrm{r}} =
3:2:1$ at the sharing radius $x_{3:2:1}=2.395$, and identifying
$\nu_{\mathrm{upp}} = \nu_{\mathrm{K}}$, we obtain the black hole
mass of Sgr\,A$^*$ in the interval
\begin{equation}
    4.29 \times 10^6 ~\mathrm{M}_{\odot} < M < 5.36 \times 10^6
    ~\mathrm{M}_{\odot}
\end{equation}
which meets the allowed black hole mass interval
(Eq.~\ref{mass-Ghez-1}) at its high mass end and is in good
agreement with the estimate (Eq.~\ref{mass-Ghez-2}).

\Section{Conclusions}
We present the possibilities for strong
resonant phenomena arising in a shared radius in the field of
black holes with spin appropriately tuned. We assume that the
strong resonances could occur and influence each other causally,
if $\nu_{\mathrm{K}}:\nu_{\theta}:\nu_{\mathrm{r}}$ is in ratio of
small integers $s:t:u$ with $s \leq 5$, when the order of the
resonances could be low enough in order to enable strong resonance
with relatively wide resonance frequency width (Landau and
Lifshitz 1976). The strong resonances with lowest values of the
frequency ratio occur for the extraordinary resonant spin
$a_{3:2:1} = 0.983$.

Due to the allowed frequency detuning in the resonant phenomena
(with maximal relative frequency detuning limited to few percent),
the strong resonant phenomena could appear in black holes with
spin values concentrated around the special resonant spin values.
For a given relative frequency detuning $\delta$, the spin scatter
differs for different special resonant spin values, being
dependent on the position of the resonant radius (and the gradient
of the frequency profiles at this radius), see Table
\ref{rozp-tab}. In the most interesting case of the extraordinary
resonant spin $a_{3:2:1}=0.983$, the spin scatter $\approx \pm
0.003$ for detuning $\delta = 0.01$ gives probably a realistic
interval of black hole spins where the strong resonant phenomena
could be observable. Of course, the value of $\delta$ (and the
spin value scatter) depend on physical conditions and resonant
phenomena in concrete sources and have to be discussed carefully
for each individual source.

There is an indication that the QPOs data observed in the Galactic
centre source Sgr\,A$^*$ imply the black hole spin close to the
extraordinary resonant value of $a \approx 0.983$, when
$\nu_{\mathrm{K}}:\nu_{\theta}:\nu_{\mathrm{r}} = $ 3\,:\,2\,:\,1
holds at the radius $x_{3:2:1} = 2.395$ and the black hole mass
can be estimated to be in the interval of $(4.8 \pm 0.5)\times
10^6 ~\mathrm{M}_{\odot}$. Therefore, it is interesting to check
concordance of the QPOs observation induced data with a variety of
the other observations of the Sgr\,A$^*$. It should be stressed
that more precise measurement of the QPOs frequencies will enable
more precise determination of the black hole mass. Moreover, we
expect NGC 5408 X-1 to be another candidate for a black hole
admitting strong resonant phenomena because of the observed
frequency set (Strohmayer \etal 2007).

It is important from the principal reason to note that if a black
hole with appropriately tuned spin $a$ admits strong resonant
phenomena with different resonances sharing the same radius of the
accretion disc, we could expect observable QPOs at resonant
frequencies even in situations when the internal disc conditions
are not convenient for starting up resonant phenomena, as
observed, \eg in Sgr\,A$^*$ source (Genzel \etal 2003).

On the other hand, the four (or five) QPOs frequencies observed in
the microquasar GRS 1915+105 can not be explained by the strong
resonances model because of specific distribution of the observed
frequency ratios (two different pairs of frequencies with 3\,:\,2
ratio). In this case, the extended resonant model including the so
called humpy oscillations induced by the ``humpy'' velocity
profile of the accretion disc, as measured in the privileged
family of locally non-rotating frames, can explain all the
observed frequencies, if the black hole parameters are fixed at
the spin $a \approx 0.9998$ and the mass $M \approx
14.4~\mathrm{M}_{\odot}$ (Stuchl{\'{\i}}k \etal 2007c), in
agreement with the spectral fit of the spin $a \approx 1$
(McClintock \etal 2006).

The conditions for strong resonant phenomena could be realized
only for black holes with high values of dimensionless spin ($a
\geq 0.75$). Therefore, the idea of strong resonant phenomena
probably could not be extended to the neutron star systems, where
we expect spin $a < 0.5$, at least if the Hartle--Thorne metric
parameters corresponding to the spin $a$ and the quadrupole
momentum $q$ are close to the quasi-Kerr values with $q\sim j^2$.
Note that for the neutron star systems probably another version of
the multiresonant idea could be realized in close agreement with
observational data, where the so called total precession resonance
model with resonant frequencies $\nu_{\mathrm{K}}$ and
$\nu_{\theta}-\nu_{\mathrm{r}}$ (related to the relativistic
precession model, Stella and Vietri 1998) is involved and realized
at different radii, as shown in Stuchl{\'{\i}}k \etal (2007e).

In the field of Kerr black holes with $a \approx 0.866$, the four
frequency set ratio 4\,:\,3\,:\,2\,:\,1 could be observed, if the
resonant conditions allow combinational frequencies to be
observable. For the spin $a \approx 0.962$ or $a \approx 0.882$,
even the five frequency set 5\,:\,4\,:\,3\,:\,2\,:\,1 could be
observable in some special circumstances. Clearly, it is not
necessary that all the resonances are realized simultaneously and
that the full five frequency set is observed at the same time.

Undoubtly, it is worth to search for such special cases of
frequency sets in observational data as these could lead to
precise determination of the black hole spin enabling a deeper
understanding of a variety of astrophysical phenomena involved in
determining behaviour of accretion discs in strong gravitational
fields. We should note that when  simple combinational frequencies
not exceeding the Keplerian frequency are allowed, the lowest
triple frequency ratio set 3\,:\,2\,:\,1 can be realized for black
holes with the extraordinary resonant spin $a \approx 0.983$, but
also for the black holes with three other values of the spin $a =
0.866, 0.882, 0.962$, if the uppermost frequencies are not
observed for some reasons.

Finally, it has to be stressed that observation of some
characteristic frequency sets with low integer ratio could be a
signature of a specific value of the black hole spin $a$ enabling
strong resonances at a fixed radius. However, generally, there
exist few values of the spin $a$ and the corresponding shared
resonance radius allowed for a given frequency ratio set
(Stuchl{\'{\i}}k \etal 2007b). Therefore, detailed analysis of the
resonance phenomena, including the resonance strength and resonant
frequency width, has to be considered in a concrete candidate
system, and it has to be further confronted with the spin
estimates coming from spectral analysis of the black hole system
as given, \eg in McClintock \etal (2006) and Middleton \etal
(2006) for GRS 1915+105, and Shafee \etal (2006) for GRO J1655-40,
the line profiles (Fabian and Miniutti 2005, Dov\v{c}iak \etal
2004, Zakharov 2003, Zakharov and Repin 2006), and the orbital
periastron precession of some stars moving in the region of
Sgr\,A$^*$ (Kraniotis 2005, 2007), in order to establish the black
hole spin. Very promising from this point of view seem to be
studies of the energy dependencies of high-frequency QPO
determining the QPO spectra at the QPO radii ({\.Z}ycki \etal
2007) as they could be credited by the strong resonance conditions
giving precisely the black hole spin and the radius where the
observed QPOs are generated.

\Acknow{This work was supported by the Czech grant MSM
4781305903.}


\begin{references}
\refitem{Abramowicz, M.A., Klu{\'z}niak, W., Stuchl{\'{\i}}k, Z.,
and T{\"{o}}r{\"{o}}k, G.}{2004}{~}{~}{in: \textit{Proceedings of
RAGtime 4/5: Workshops on black holes and neutron stars}, Opava,
14--16/13--15 October 2002/2003, ed. S.~Hled{\'{\i}}k and
  Z.~Stuchl{\'{\i}}k (Opava: Silesian University in Opava), pp.
  1--23}

\refitem{Aliev, A.N., and Galtsov, D.V.}{1981}{General Relativity
and Gravitation}{13}{899}

\refitem{Aschenbach, B.}{2004}{\AA}{425}{1075,
arXiv:astro-ph/0406545v1}

\refitem{Aschenbach, B.}{2006}{Chinese Journal of Astronomy and
Astrophysics}{6}{221, arXiv:astro-ph/0603193v1}

\refitem{Aschenbach, B., Grosso, N., Porquet, D., and Predehl,
P.}{2004}{\AA}{417}{71, arXiv:astro-ph/0401589v2}

\refitem{Barret, D., Olive, J.-F., and Miller,
M.C.}{2005}{\MNRAS}{361}{855, arXiv:astro-ph/0505402v1}

\refitem{Belloni, T., M{\'e}ndez, M., and Homan,
J.}{2005}{\AA}{437}{209, arXiv:astro-ph/0501186v2}

\refitem{Belloni, T., M{\'e}ndez, M., and Homan,
J.}{2007}{\MNRAS}{376}{1133, arXiv:astro-ph/0702157v1}

\refitem{Belloni, T., M{\'e}ndez, M., and
S\'{a}nchez-Fern\'{a}ndez, C.}{2001}{\AA}{372}{551,
arXiv:astro-ph/0104019v1}

\refitem{Dov\v{c}iak, M., Karas, V., Martocchia, A., Matt, G., and
Yaqoob, T.}{2004}{~}{~}{in: \textit{Proceedings of RAGtime 4/5:
Workshops on black holes and neutron stars}, Opava, 14--16/13--15
October 2002/2003, ed. S.~Hled{\'{\i}}k and
  Z.~Stuchl{\'{\i}}k (Opava: Silesian University in Opava),
  pp. 33--73, arXiv:astro-ph/0407330v1}

\refitem{Fabian, A.C., and Miniutti, G.}{2005}{~}{~}{``Kerr
Spacetime: Rotating Black Holes in General Relativity''
(Cambridge: Cambridge University Press), eprint arXiv:astro-ph/0507409v1
is a part of this book}

\refitem{Genzel, R., Sch{\"o}del, R., Ott, T., Eisenhauer, F.,
Hofmann, R., Lehnert, M., Eckart, A., Alexander, T., Sternberg,
A., Lenzen, R., Cl{\'e}net, Y., Lacombe, F., Rouan, D., Renzini,
A., and Tacconi-Garman, L.E.}{2003}{\ApJ}{594}{812,
arXiv:astro-ph/0305423v1}

\refitem{Ghez, A.M., Salim, S., Weinberg, N.N., Lu, J.R., Do, T.,
Dunn, J.K., Matthews, K., Morris, M., Yelda, S., Becklin, E.E.,
Kremenek, T., Milosavljevic, M., and Naiman,
J.}{2008}{\ApJ}{689}{1044, arXiv:0808.2870v1 [astro-ph]}

\refitem{Hartle, J.B., and Thorne, K.S.}{1968}{\ApJ}{153}{807}

\refitem{Kato, S., Fukue, J., and Mineshige,
S.}{1998}{~}{~}{``Black-hole accretion disks'', ed. S.~Kato,
J.~Fukue, and S.~Mineshige (Kyoto, Japan: Kyoto University Press)}

\refitem{Klu{\'z}niak, W., and Abramowicz, M.A.}{2000}{Phys. Rev.
Lett.}{~}{submitted, arXiv:astro-ph/0105057v1}

\refitem{Kraniotis, G.V.}{2005}{Classical Quantum
Gravity}{22}{4391, arXiv:gr-qc/0507056v3}

\refitem{Kraniotis, G.V.}{2007}{Classical Quantum
Gravity}{24}{1775, arXiv:gr-qc/0602056v4}

\refitem{Lachowicz, P., Czerny, B., and Abramowicz,
M.A.}{2006}{\MNRAS}{~}{submitted, arXiv:astro-ph/0607594v1}

\refitem{Landau, L.D., and Lifshitz, E.M.}{1976}{~}{~}{``Course of
Theoretical Physics'', Vol.~I, Mechanics, 3rd edn. (Oxford:
Elsevier Butterworth-Heinemann)}

\refitem{McClintock, J.E., and Remillard,
R.A.}{2004}{~}{~}{``Compact Stellar X-Ray Sources'', ed. W.~H.~G.
Lewin and M.~van~der Klis (Cambridge: Cambridge University Press)}

\refitem{McClintock, J.E., Shafee, R., Narayan, R., Remillard,
R.A., Davis, S.W., and
  Li, L.-X.}{2006}{\ApJ}{652}{518, arXiv:astro-ph/0606076v2}

\refitem{Middleton, M., Done, C., Gierli{\'n}ski, M., and Davis,
S.W.}{2006}{\MNRAS}{373}{1004, arXiv:astro-ph/0601540v2}

\refitem{Nayfeh, A.H., and Mook, D.T.}{1979}{~}{~}{``Nonlinear
Oscillations'', New York: Wiley}

\refitem{Novikov, I.D., and Thorne, K.S.}{1973}{~}{~}{``Black
Holes'', ed. C.~D. Witt and B.~S.~D. Witt (New
York--London--Paris: Gordon and Breach), p. 343}

\refitem{Perez, C.A., Silbergleit, A.S., Wagoner, R.V., and Lehr,
D.E.}{1997}{\ApJ}{476}{589, arXiv:astro-ph/9601146v2}

\refitem{Reid, M.J.}{2008}{Internat. J. Modern Phys.
D}{~}{accepted, arXiv:0808.2624v1 [astro-ph]}

\refitem{Remillard, R.A.}{2005}{\AN}{326}{804,
arXiv:astro-ph/0510699v1}

\refitem{Remillard, R.A., and McClintock, J.E.}{2006}{Annual
Review of Astronomy and Astrophysics}{44}{49,
arXiv:astro-ph/0606352v1}

\refitem{Rezzolla, L.}{2004a}{~}{~}{``X-ray Timing 2003: Rossi and
Beyond'', ed. P.~Karet, F.~K. Lamb, and J.~H. Swank, Vol. 714
(Melville: NY: American Institut of Physics), p. 36}

\refitem{Rezzolla, L.}{2004b}{~}{~}{in: \textit{Proceedings of
RAGtime 4/5: Workshops on black holes and neutron stars}, Opava,
14--16/13--15 October 2002/2003, ed. S.~Hled{\'{\i}}k and
Z.~Stuchl{\'{\i}}k (Opava: Silesian University in Opava), pp.
151--165}

\refitem{Rezzolla, L., Yoshida, S., Maccarone, T.J., and Zanotti,
O.}{2003}{\MNRAS}{344}{L37, arXiv:astro-ph/0307487v1}

\refitem{Schnittman, J.D., and Rezzolla,
L.}{2006}{\ApJ}{637}{L113, arXiv:astro-ph/0506702v1}

\refitem{Shafee, R., McClintock, J.E., Narayan, R., Davis, S.W.,
Li, L.-X., and Remillard, R.A.}{2006}{\ApJ}{636}{L113,
arXiv:astro-ph/0508302v2}

\refitem{{\v S}r{\'{a}}mkov{\'{a}}, E.}{2005}{\AN}{326}{835}

\refitem{Stella, L., and Vietri, M.}{1998}{Astrophys. J.
Lett.}{492}{L59, arXiv:astro-ph/9709085v1}

\refitem{Strohmayer, T.E.}{2001}{\ApJ}{552}{L49}

\refitem{Strohmayer, T.E., Mushotzky, R., Winter, L., Soria, R.,
Uttley, P., and Cropper, M.}{2007}{\ApJ}{660}{580,
arXiv:astro-ph/0701390v1}

\refitem{Stuchl{\'{\i}}k, Z., Slan{\'{y}}, P., and
T{\"{o}}r{\"{o}}k, G.}{2004}{~}{~}{in: \textit{Proceedings of
RAGtime 4/5: Workshops on black holes and neutron stars}, Opava,
14--16/13--15 October 2002/2003, ed. S.~Hled{\'{\i}}k and
  Z.~Stuchl{\'{\i}}k (Opava: Silesian University in Opava),
  pp. 239--256}

\refitem{Stuchl{\'{\i}}k, Z., and Hled{\'{\i}}k,
S.}{2005}{~}{~}{in: \textit{Proceedings of RAGtime 6/7:
  Workshops on black holes and neutron stars}, Opava, 16--18/18--20 September
  2004/2005, ed. S.~Hled{\'{\i}}k and Z.~Stuchl{\'{\i}}k (Opava: Silesian
  University in Opava), pp. 189--208}

\refitem{Stuchl{\'{\i}}k, Z., and T{\"{o}}r{\"{o}}k,
G.}{2005}{~}{~}{in: \textit{Proceedings of RAGtime
  6/7: Workshops on black holes and neutron stars}, Opava, 16--18/18--20
  September 2004/2005, ed. S.~Hled{\'{\i}}k and Z.~Stuchl{\'{\i}}k (Opava:
  Silesian University in Opava), pp. 253--263}

\refitem{Stuchl{\'{\i}}k, Z., Slan{\'{y}}, P., T{\"{o}}r{\"{o}}k,
G., and Abramowicz, M.A.}{2005}{Phys. Rev. D}{71}{024037}

\refitem{Stuchl{\'{\i}}k, Z., Slan{\'{y}}, P., and
T{\"{o}}r{\"{o}}k, G.}{2006}{~}{~}{in: \textit{Proceedings of the
VI Microquasar Workshop: Microquasars and Beyond}, Como,
  Italy, 18--22 September 2006, p. 95, arXiv:astro-ph/0612439v1}

\refitem{Stuchl{\'{\i}}k, Z., Konar, S., Miller, J.C., and
Hled{\'{\i}}k, S.}{2007a}{~}{~}{in: \textit{Proceedings of RAGtime
8/9: Workshops on black holes and neutron stars}, Opava, Hradec
nad Moravic\'{i}, 15--19/19--21 September 2006/2007, ed.
S.~Hled{\'{\i}}k and Z.~Stuchl{\'{\i}}k (Opava: Silesian
  University in Opava), pp. 293--322}

\refitem{Stuchl{\'{\i}}k, Z., Kotrlov{\'{a}}, A., and
T{\"{o}}r{\"{o}}k, G.}{2007b}{~}{~}{in: \textit{Proceedings of
RAGtime 8/9: Workshops on black holes and neutron stars}, Opava,
Hradec nad Moravic\'{i}, 15--19/19--21 September 2006/2007, ed.
S.~Hled{\'{\i}}k and Z.~Stuchl{\'{\i}}k (Opava: Silesian
  University in Opava), pp. 363--416}

\refitem{Stuchl{\'{\i}}k, Z., Slan{\'{y}}, P., and
T{\"{o}}r{\"{o}}k, G.}{2007c}{\AA}{463}{807}

\refitem{Stuchl{\'{\i}}k, Z., Slan{\'{y}}, P., and
T{\"{o}}r{\"{o}}k, G.}{2007d}{\AA}{470}{401, arXiv:0704.1252v2
[astro-ph]}

\refitem{Stuchl{\'{\i}}k, Z., T{\"{o}}r{\"{o}}k, G., and Bakala,
P.}{2007e}{\AA}{~}{submitted, arXiv:0704.2318v2 [astro-ph]}

\refitem{Stuchl{\'{\i}}k, Z., Konar, S., Miller, J.C., and
Hled{\'{\i}}k, S.}{2008}{\AA}{489}{963, arXiv:0808.3641v1
[astro-ph]}

\refitem{T{\"{o}}r{\"{o}}k, G.}{2005}{\AA}{440}{1,
arXiv:astro-ph/0412500v1}

\refitem{T{\"{o}}r{\"{o}}k, G., Abramowicz, M.A., Klu{\'z}niak,
W., and Stuchl{\'{\i}}k, Z.}{2005}{\AA}{436}{1}

\refitem{T{\"{o}}r{\"{o}}k, G., and Stuchl{\'{\i}}k,
Z.}{2005a}{~}{~}{in: \textit{Proceedings of RAGtime 6/7: Workshops
on black holes and neutron stars}, Opava, 16--18/18--20 September
2004/2005, ed. S.~Hled{\'{\i}}k and Z.~Stuchl{\'{\i}}k (Opava:
Silesian University in Opava), pp. 315--338}

\refitem{T{\"{o}}r{\"{o}}k, G., and Stuchl{\'{\i}}k,
Z.}{2005b}{\AA}{437}{775, arXiv:astro-ph/0502127v1}

\refitem{van~der Klis, M.}{2000}{Annual Review of Astronomy and
Astrophysics}{38}{717, arXiv:astro-ph/0001167v1}

\refitem{van~der Klis, M.}{2006}{~}{~}{``Compact Stellar X-Ray
Sources'', ed. W.~H.~G. Lewin and M.~van~der Klis (Cambridge:
Cambridge University Press), pp. 39--112}

\refitem{Zakharov, A.F.}{2003}{Publications of the Astronomical
Observatory of Belgrade}{76}{147}

\refitem{Zakharov, A.F., and Repin, S.V.}{2006}{New
Astronomy}{11}{405, arXiv:astro-ph/0510548v1}

\refitem{{\.Z}ycki, P.T., Nied{\'z}wiecki, A., and Sobolewska,
M.A.}{2007}{\MNRAS}{379}{123, arXiv:0704.3394v1 [astro-ph]}
\end{references}
\end{document}